\documentclass[brazilian,twocolumn,prl,superscriptaddress,amsmath,amssymb,showpacs,floatfix,preprintnumbers]{revtex4-2}
\usepackage[latin9,utf8]{inputenc}
\usepackage{float}
\usepackage{textcomp}
\usepackage{amsmath}
\usepackage{graphicx}
\usepackage{url}
\usepackage{grffile}
\usepackage{bbold}
\usepackage{dcolumn}
\usepackage{color}
\DeclareGraphicsExtensions{.png .jpg .pdf}
\usepackage{hyperref}
\hypersetup{
     colorlinks = true,
     linkcolor = blue,
     anchorcolor = blue,
     citecolor = blue,
     filecolor = blue,
     urlcolor = blue
     }
\usepackage{braket}
\usepackage{physics}
\usepackage{array}
\usepackage{booktabs}
\usepackage{soul}

\makeatother

\begin{document}


\title{Distinctive $g$-factor of moiré-confined excitons in van der Waals heterostructures}

\author{Y. Galvão Gobato}\email{yara@df.ufscar.br}

\author{C. Serati de Brito}
\affiliation{Physics Department, Federal University of São Carlos (UFSCAR)-Brazil}

\author{A. Chaves}\email{andrey@fisica.ufc.br}
\affiliation{Universidade Federal do Cear\'a, Departamento de F\'{\i}sica, 60455-760 Fortaleza, Cear\'a, Brazil}
\affiliation{Department of Physics, University of Antwerp, Groenenborgerlaan 171, B-2020 Antwerp, Belgium}

\author{M. A. Prosnikov}
\affiliation{High Field Magnet Laboratory (HFML–EMFL), Radboud University, 6525 ED, Nijmegen, The Netherlands}

\author{T. Wo\'zniak}
\affiliation{Department of Semiconductor Materials Engineering, Wroc\l aw University of Science and Technology, 50-370 Wroc\l aw, Poland}

\author{Shi Guo}
\affiliation{Centre for Graphene Science, College of Engineering, Mathematics and Physical Sciences, University of Exeter, Exeter, EX4 4QF}

\author{Ingrid D. Barcelos}
\affiliation{Brazilian Synchrotron Light Laboratory (LNLS), Brazilian Center for Research in Energy and Materials (CNPEM), 13083-970 Campinas, São Paulo, Brazil}

\author{M. V. Milo\v{s}evi\'c}
\affiliation{Department of Physics, University of Antwerp, Groenenborgerlaan 171, B-2020 Antwerp, Belgium}
\affiliation{NANOlab Center of Excellence, University of Antwerp, Belgium}

\author{F. Withers}
\affiliation{Centre for Graphene Science, College of Engineering, Mathematics and Physical Sciences, University of Exeter, Exeter, EX4 4QF}

\author{P. C. M. Christianen}
\affiliation{High Field Magnet Laboratory (HFML–EMFL), Radboud University, 6525 ED, Nijmegen, The Netherlands}
 
\begin{abstract}
We investigated experimentally the valley Zeeman splitting of excitonic peaks in the photoluminescence (PL) spectra of high-quality hBN/WS$_2$/MoSe$_2$/hBN  heterostructures at near-zero twist angles under perpendicular magnetic fields up to 20 T. We identify two neutral exciton peaks in the PL spectra: the lower energy one exhibits a reduced $g$-factor relative to that of the higher energy peak, and much lower than the recently reported values for interlayer excitons in other van der Waals (vdW) heterostructures. We provide evidence that such a discernible $g$-factor stems from the spatial confinement of the exciton in the potential landscape created by the moiré pattern, due to lattice mismatch and/or inter-layer twist in heterobilayers. This renders magneto-$\mu$PL an important tool to reach deeper understanding of the effect of moiré patterns on excitonic confinement in vdW heterostructures.
\end{abstract}

\maketitle

Recent years have witnessed enormous interest in the optical properties of heterostructures based on monolayer transition-metal dichalcogenides (TMDs) \cite{geim2013van, novoselov20162d}. Monolayer TMDs exhibit a direct band gap at two inequivalent $\pm$K valleys and enhanced excitonic effects due to strong out-of-plane quantum confinement, large in-plane carrier effective mass and significantly reduced Coulomb screening \cite{Mak2010,Splendiani2010,Xu2014,Xiao2012}. Large spin-orbit coupling and lack of inversion symmetry lift the spin degeneracy of the electron and hole states and therefore lock the spin and valley degrees of freedom \cite{Xiao2012,Mak2012,Zeng2012,Schaibley2016}. As a consequence, the $\pm K$ valleys can be individually addressed using right- ($\sigma_+$) and left-handed ($\sigma_-$) circularly polarized light \cite{ishii2019optical}. Under perpendicular external magnetic fields, valley Zeeman effects and magnetic-field-induced valley polarization can be observed \cite{Aivazian2015,Li2014,Srivastava2015,Macneill2015,Wang2015,Mitioglu2015,Stier2016,Plechinger2016}.   
 

TMD heterostructures with stacking angles near either 0$^\circ$ (AA stacking) or 60$^\circ$ (AB stacking) exhibit pronounced emission from both inter-layer and intra-layer excitons \cite{Rivera2016, Nayak2017,Nagler2017,Forg2021}. They both present large exciton binding energies ($>$100 meV) and are highly robust against dissociation under applied electric fields, which opens up new opportunities to investigate valley–spin optoelectronics \cite{Rivera2018,Tartakovskii2020}. The lattice mismatch between the materials that compose the heterostructure, combined with small inter-layer twist angles, lead to a moiré pattern consisting of regions where the crystal exhibits stacking registries that are locally different along the materials plane \cite{kuwabara1990anomalous}. Recent experiments have shown evidence that such different local crystal properties in the moiré pattern effectively produce a potential landscape strong enough to localize both inter- and intra-layer exciton wave functions and thus produce quantized exciton states \cite{tran2019evidence}, which brings the possibility of using moiré-confined excitons as single-photon emitters and quantum simulators \cite{baek2020highly, huang2022excitons}.

It was previously shown that the effective $g$-factor of inter-layer excitons (iXs) depends on the twist angle between the layers of van der Waals heterostructures (vdWHSs) \cite{seyler2019signatures}. Several experimental studies of MoSe$_2$/WSe$_2$ heterostructures have shown an effective $g$-factor of iXs of $g \approx +6$ for twist angles near 0$^\circ$, while $g \approx -15$ was found for twist angles around 60$^\circ$ \cite{Nagler2017,seyler2019signatures,forg2021moire}. These values are very different from reported $g$-factors of intra-layer excitons in individual monolayer TMDs ($g\approx -4$) \cite{Li2014,Macneill2015,Wang2015,seyler2019signatures}. In fact, the value of effective exciton g-factor results from spin and orbital angular momenta of electron and hole bands, while its sign is governed by optical selection rules \cite{wozniak2020exciton, deilmann2020ab}. As a consequence, it is significantly affected by the materials composition, inter-layer exciton hybridization, and even by the spread of the exciton wave function \cite{chen2019luminescent}. In the presence of the effective confining potential produced by the moiré pattern, the confined exciton wave function in reciprocal space is expected to populate regions away from the K-point edge of the Brillouin zone, thus effectively modifying the exciton $g$-factor, which can in turn be used to probe the moiré exciton confinement.

In this work we experimentally analyze the exciton $g$-factors in a MoSe$_2$/WS$_2$ heterobilayer with near-zero twisting angle. Due to the relatively large lattice mismatch between these two TMDs, strong moiré exciton confinement is expected, even for a negligible inter-layer twist as in our samples. Our magneto-photoluminescence (magneto-$\mu$PL) data reveal the existence of two neutral exciton peaks, at different energies. The effective $g$-factor of the lower-energy exciton is consistently lower as compared to either the perfectly charge-separated inter-layer excitons observed in MoSe$_2$/WSe$_2$ (and other) heterobilayers, or the intra-layer excitons either in an individual MoSe$_2$ monolayer or in MoSe$_2$/WS$_2$ heterobilayers. 
We argue that the low energy PL peak is due to an exciton whose center-of-mass motion is confined by the moiré potential landscape, as evidenced by its energy separation from the peak identified as the MoSe$_2$ intra-layer exciton (with an experimentally validated $g$-factor). 

The monolayers of MoSe$_2$ and WS$_2$ were obtained by mechanical exfoliation from bulk single crystals from HQ Graphene. The heterostructures were fabricated by an all-dry transfer technique and consist of a monolayer of MoSe$_2$ on a monolayer of WS$_2$, sandwiched between two thin sheets of hexagonal boron nitride (hBN). The small stacking angle $\approx$ 0$^\circ$ is obtained by controlling the orientation of the edges of the cleaved flakes and the crystal direction. For $\mu$PL measurements, the sample was placed on a x–y–z piezoelectric Attocube stage and cooled down to 4.2 K in a cryostat placed inside a 30 T resistive magnet. The $\mu$PL measurements were performed using a continuous-wave Millennia Spectra Physics laser excitation with a photon energy of 2.33 eV. The laser was focused on the flake by an 40$\times$ Attocube objective (NA = 0.55) with a spot size of about 4 $\mu$m. The $\mu$PL signal was collected with appropriate optics in backscattering configuration and measured by a Princeton Instruments Acton spectrometer connected with a liquid-nitrogen cooled charge-coupled device. The polarization-resolved excitation and $\mu$PL collection were performed using  quarter-wave plates and linear polarizers. The sample was mounted with its normal parallel to the magnetic field direction (Faraday configuration) and kept at a temperature of 4.2 K. We have measured the $\mu$PL at several laser positions on the heterostructure region, which allowed us to verify the consistency of our findings. All results reported in this Letter are taken at a single location in the sample, while results for other locations are made available in the Supplemental Material (SM) \cite{SM}.

\begin{figure}[!t]
\centering{\includegraphics[width=1\columnwidth]{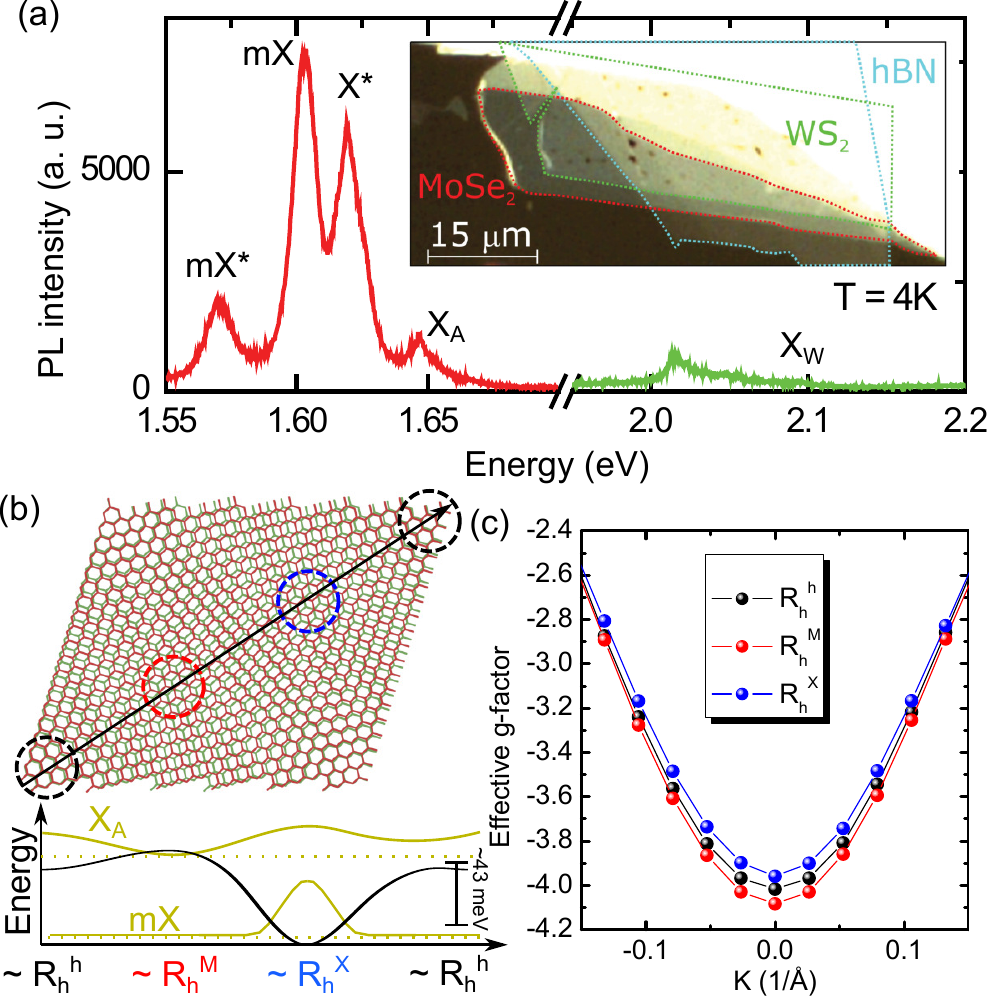}}
\caption{(Color online) (a) Typical PL spectrum of the MoSe$_2$/WS$_2$ vdWHS at 4 K. The inset shows the optical image of the sample. The intersection of red (MoSe$_2$) and green (WS$_2$) delimited areas highlights the vdWHS region. PL of the monolayer WS$_2$ region of the sample (green data points) shows features at higher energies, identified as the monolayer exciton X$_W$ peak for this material, whereas the lower energy features are identified as the free (moiré-confined) MoSe$_2$ intra-layer exciton X$_A$ (mX) and trion X$^*$ (mX$^*$). (b) Sketch of the crystal structure of a MoSe$_2$/WS$_2$ vdWHS with a twist angle $\approx$ 0. Local stacking registries $R_h^h$, $R_h^M$ and $R_h^X$, highlighted respectively by black, red, and blue circles, form a periodic moiré pattern. The quasi-particle energy gap is locally different in each of these regions, thus producing a moiré potential landscape for excitons. This is illustrated by the black line in the bottom panel, which depicts the quasi-particle gap along the arrow drawn in the top panel. The mX and X$_A$ peaks in the PL spectrum originate respectively from confined and (nearly) free excitons in the moiré pattern, whose wave functions are represented by yellow curves in the bottom panel. (c) Effective $g$-factor dispersion for the intra-layer MoSe$_2$ exciton in the vicinity of the K-point, as obtained by DFT calculations assuming the three stacking registries identified in (b). }\label{fig.sketch}
\end{figure}

As shown in Fig. \ref{fig.sketch}(a), the PL spectrum of the MoSe$_2$/WS$_2$ vdWHS, whose optical microscopy image is shown in the inset, measured in absence of an applied magnetic field, exhibits two peaks associated to the neutral excitons, labeled mX (1.603 eV) and X$_A$ (1.646 eV), interspersed by trion peaks mX$^*$ (1.570 eV) and X$^*$ (1.619 eV). These results are very similar to previous experimental results reported for MoSe$_2$/WS$_2$ heterobilayers \cite{alexeev2019resonantly, zhang2020twist}. We note the energy separation between the mX and X$_A$ PL peak positions of $\approx$ 43 meV. The energy separation between X$_A$ and X$^*$ is 27 meV, whereas between mX$^*$ and mX peaks a splitting of 33 meV was observed. The energy separations are consistent with the identification of mX$^*$ and X$^*$ as trion emissions. Our laser power dependence study of all observed PL bands at 4 K (Fig. S3 in the SM \cite{SM}) provides further support to this interpretation of the excitonic nature of different PL peaks. We thus correlate the experimentally observed mX peak to a moiré-confined exciton state, whereas the X$_A$ peak is interpreted as the emission from an unconfined intra-layer MoSe$_2$ exciton \cite{tang2021tuning,alexeev2019resonantly, zhang2020twist}, as will be detailed further below.


In order to understand the origin of the two neutral exciton peaks consistently observed in our experiments, let us consider a van der Waals stacked MoSe$_2$/WS$_2$ hetero-bilayer with inter-layer twist angle $\approx 0^{\circ}$, as sketched in Fig. \ref{fig.sketch}(b). The periodic moiré pattern created by the lattice mismatch between the layers exhibits regions with local stacking registries identified as $R_h^h$, $R_h^M$ and $R_h^X$, marked as black, red, and blue circles, respectively. Since the quasi-particle gap of each of these registries is different, one expects these regions to form a periodic landscape of potential wells and barriers for the exciton \cite{tran2019evidence}, as illustrated by the black line in the bottom panel of Fig. \ref{fig.sketch}(b). The quasi-particle gaps obtained from ab initio calculations with hybrid functional (HSE06) in a MoSe$_2$/WS$_2$ vdWHS with $R_h^h$, $R_h^M$ and $R_h^X$ are 1.830 eV, 1.831, and 1.816 eV, respectively. All three cases exhibit practically the same effective mass and, consequently, the same binding energy 0.230 eV, thus explaining the experimentally observed peaks around 1.6 eV. In the $R_h^X$ case, however, the theoretically predicted exciton energy is at least 15 meV lower than those of $R_h^h$ and $R_h^M$, thus working as a confining well for the exciton center-of-mass. Since the inter-atomic distances in the actual vdWHS crystal are likely different from those in our simulations for pure $R_h^h$, $R_h^M$ and $R_h^X$ crystals, which were artificially strained to produce commensurate lattices, and our calculations for local quasi-particle gaps were made without GW corrections, the actual moiré confinement potential in the MoSe$_2$/WS$_2$ vdWHS is likely to be significantly stronger than the $\approx $15 meV predicted here.  

The valence band edges of the isolated monolayers are separated by $\approx$ 750 meV \cite{haastrup2018computational}, which restricts the holes in the lowest energy exciton states to the MoSe$_2$ layer. Conversely, the conduction band offset between the individual monolayers is small: for instance, Refs. [\onlinecite{alexeev2019resonantly}] and [\onlinecite{zhang2020twist}] suggest conduction band offsets of 13 meV and 5.8 meV, respectively, with lowest energy electrons also in the MoSe$_2$ layer. Ab initio calculations in Ref. [\onlinecite{haastrup2018computational}] give somewhat higher values for this band offset, but all results in the literature point towards a type-I band alignment for this combination of materials.\cite{tang2022dielectric} Therefore, in principle, the lowest energy exciton state in this vdW HS must have a dominant intra-layer MoSe$_2$ component, although, due to such low conduction band offset, hybrid states with an inter-layer component, where electrons partially populate the WS$_2$ layer as well, may also be possible.

The interplay between the effects of the moiré confinement potential and inter-layer hybridization in MoSe$_2$/WS$_2$ heterostructures has been thoroughly investigated in Ref. [\onlinecite{alexeev2019resonantly}], where PL spectra were observed with the same four peaks reported here, considering a similar interpretation for X$_A$ and mX, but emphasizing a possibility of inter-layer hybridization in the latter. On the other hand, the exciton Stark shift measurements reported in Ref. [\onlinecite{tang2021tuning}] provided strong evidence that the $\approx$ 1.60 eV signal (mX) in this system does not exhibit a significant out-of-plane dipole moment (i.e., high contribution from an inter-layer component), but rather stems from a virtually pure intra-layer exciton, fully consistent with the type-I band alignment. With that in mind, we will neglect the inter-layer contribution to the mX peak and focus primarily on the in-plane moiré confinement of the intra-layer exciton and how it affects the effective exciton $g$-factor for this peak under applied magnetic field. 

\begin{figure}[!t]
\centering{\includegraphics[width=1\columnwidth]{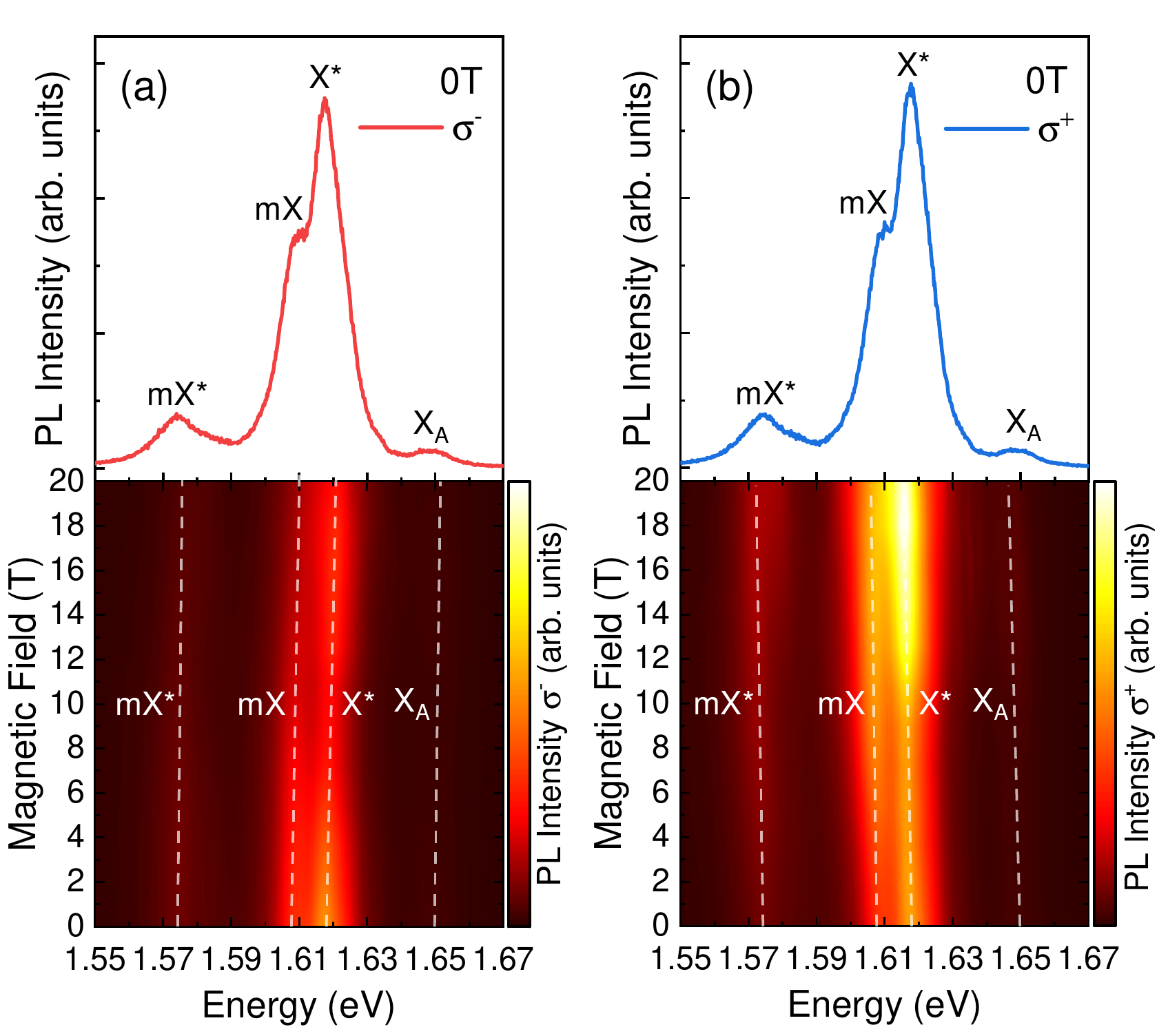}}
\caption{(Color online) Typical spectra at 0 T (top panels) for the MoSe$_2$/WS$_2$ vdWHS at 4 K and the corresponding color-code map of the PL intensity versus energy and magnetic field (bottom panels), for (a) $\sigma_-$ and (b) $\sigma_+$ light polarizations.}\label{fig.colormap}
\end{figure}

In the presence of a perpendicularly applied magnetic field $\vec B = B \hat z$, the energy of each exciton state shifts as
\begin{equation}
E_{i}(B) = E_{i}(0) \pm g_{i}\mu_B B/2,
\label{eq.magneticshifts}
\end{equation}
where $\mu_B$ is the Bohr magneton, $B$ the magnetic field strength, $i$ = mX, mX$^*$, X$^*$ or X$_A$, the $\pm$ sign depends on the valley degree of freedom, associated with the circular polarization $\sigma_{\pm}$, and the linear term corresponds to the Zeeman shift, with an effective Landé $g$-factor $g_{i}$. The quadratic (diamagnetic) shifts for each exciton state are an order of magnitude lower than the Zeeman shifts and are therefore neglected. These $g$-factors are directly related to the angular momenta of the valence and conduction band states involved in the exciton transition. Although a survey of such angular momenta values for several monolayer TMDs, along with a thorough description of the calculation method, can be found in Ref. [\onlinecite{wozniak2020exciton}], we briefly describe the theoretical procedure to obtain effective exciton $g$-factors from density functional theory calculations in the SM \cite{SM}, for the sake of completeness. 

Results in Fig. \ref{fig.sketch}(c) demonstrate that the angular momenta of the conduction and valence band states in a MoSe$_2$/WS$_2$ vdWHS with three different stacking registries observed for $\approx$ 0$^\circ$ twist angle in Fig. \ref{fig.sketch}(b) are not significantly different from that of the isolated MoSe$_2$ monolayer, which is $g \approx - 4$ \cite{wozniak2020exciton}. On the other hand, they are strongly reduced in modulus for exciton momenta away from the exciton band edge, ranging from $-4.1 < g < -3.9$ exactly at the K point, down to $g \approx -2.6$ for momenta 0.15 \AA\,$^{-1}$ away from this point. This demonstrates that if the exciton center-of-mass momentum is effectively non-zero, the exciton $g$-factor is expected to be reduced. In fact, the momentum of a ground-state exciton that is quantum confined by an external potential is naturally non-zero. The moiré potential illustrated in the bottom panel of Fig. \ref{fig.sketch}(b), with minima at the $R_h^X$ regions, suggests that the exciton center-of-mass is effectively quantum confined to these regions, which allows one to predict a significant reduction in the $g$-factors for moiré confined exciton states. Wave functions of shallow (X$_A$) and (mX) confined exciton states are depicted by yellow lines in the bottom panel of Fig. \ref{fig.sketch}(b).

Figure \ref{fig.colormap} shows the magnetic field dependence of the PL spectrum for the left ($\sigma_-$, a) and right ($\sigma_+$, b) circularly polarized emissions in a magnetic field up to 20 T for another laser position as compared to Figure 1. With increasing magnetic field, we observe that all PL peak energies shift monotonically. A red/blue shift of the circularly polarized $\sigma_+$/$\sigma_-$ PL peak positions was observed with increasing magnetic field, which clearly evidences the regular valley Zeeman splitting effect \cite{Aivazian2015,Li2014,Srivastava2015,Macneill2015,Wang2015,Mitioglu2015,Stier2016,Plechinger2016}. The perpendicular magnetic field removes the valley degeneracy and splits the PL peaks with Zeeman energies that depend linearly on the field. Furthermore, the $\sigma_+ (\sigma_-)$ polarized PL intensities increase (decrease) significantly with field evidenced by the brighter(darker) colors in Fig. \ref{fig.colormap}(b). Similar behavior was observed for $\mu$PL data obtained at different laser positions of the heterostructure (see Figs. S4 and S5 in the SM \cite{SM}).

\begin{figure}[!t]
\centering{\includegraphics[width=1\columnwidth]{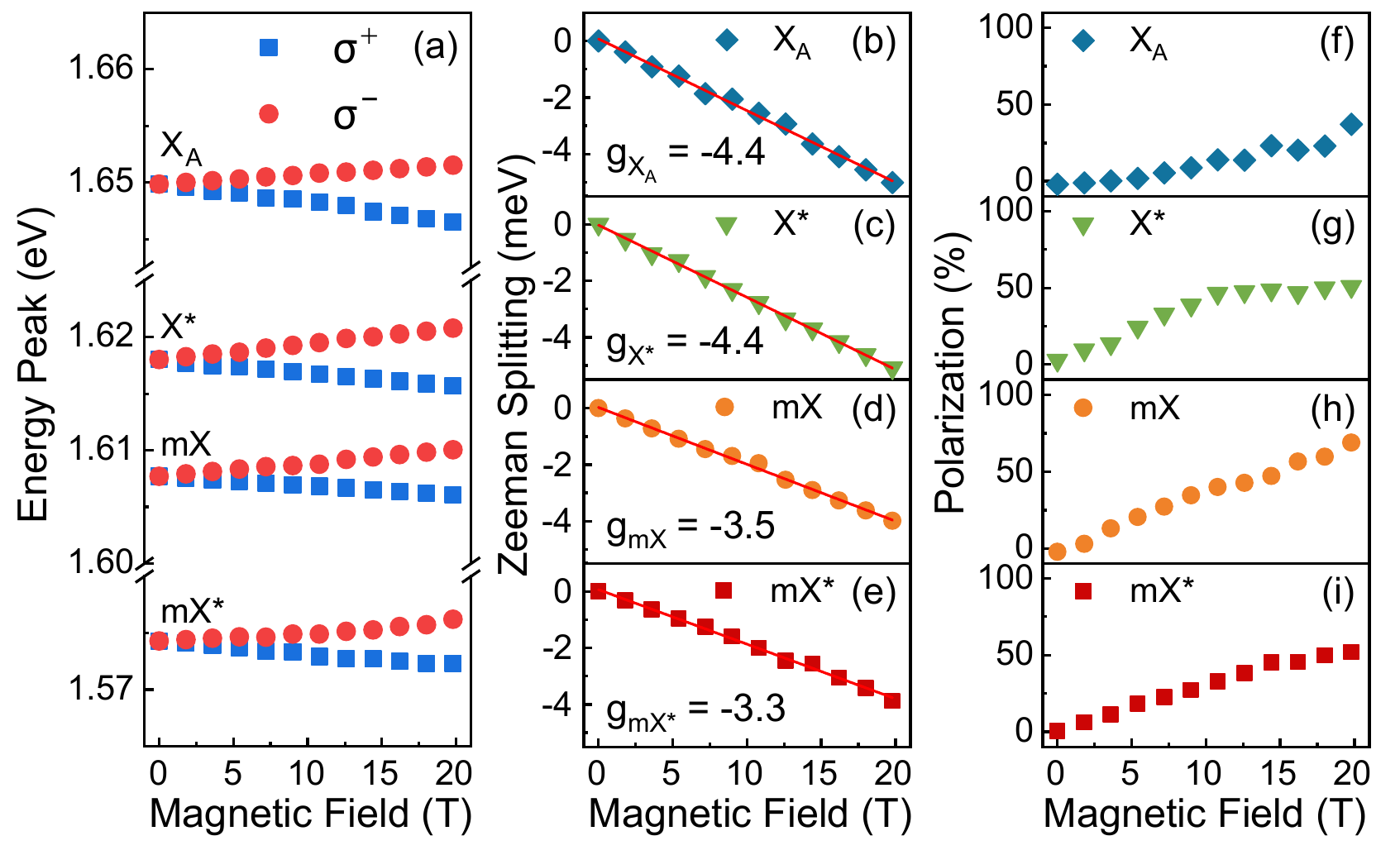}}
\caption{(Color online) Effective $g$-factors of the exciton peaks in  hBN encapsulated MoSe$_2$/WS$_2$ vdWHS. (a) Magnetic field dependence of the PL peak energies for $\sigma_+$ and $\sigma_-$ polarization. (b)-(e) Zeeman splitting of each PL peak as a function of magnetic field. Solid lines are linear fits to the data. (f)-(i) Circular polarization degree as a function of the magnetic field.}\label{fig.gfactorsexp}
\end{figure}

We have determined the peak energies of the mX, mX$^*$, X$_A$ and X$^*$ emission lines as a function of magnetic field by fitting each PL peak with a Voigt line shape. The resulting field dependence of peak energies is plotted in Fig. \ref{fig.gfactorsexp}(a). Measurements at all the positions across the heterostructure exhibit a valley splitting that clearly follows a linear dependence with magnetic field. The effective $g$-factors for all emission peaks are obtained by fitting the experimental Zeeman shifts to $\Delta E_i = E_i^{\sigma_+}-E_i^{\sigma_-} = g_i \mu_B B$ (Fig. \ref{fig.gfactorsexp}(b-e)). This yields $g$-factors $\approx$-4.4 for both the X$_A$ and X$^*$ PL peaks. Lower $g$-factors are observed for the mX$^*$ trion and mX exciton, with values $g_{mX^*} = -3.3$ and $g_{mX}=-3.5$, respectively. These values are also lower than the effective $g$-factors of excitons in monolayer MoSe$_2$, with $g_{A} \approx -4$ in previous works \cite{Koperski2019}. In fact, we consistently observe reduced effective $g$-factors for the mX and mX$^*$ PL peaks for all measured laser positions within the heterostructure region. 

As previously discussed, moiré confined excitons are expected to exhibit reduced $g$-factors. We therefore address the X$_A$ peak as either free or shallow-confined exciton states in the higher energy $R_h^h$ and $R_h^M$ regions, whereas mX is interpreted as a  moiré-confined exciton state at the $R_h^X$ region, with lower energy and, consequently, non-zero center-of-mass momentum. This interpretation of X$_A$ and mX matches the findings of earlier works \cite{alexeev2019resonantly}, except for our assumption of weak inter-layer hybridization in mX, in accordance with Ref. [\onlinecite{tang2021tuning}]. The observed $g$-factor reduction evidences the effect of moiré confinement on the exciton effective angular momentum, which is the main finding of this work. 

Let us compare the experimentally observed $\approx$ 22\% difference between $g_A$ and $g_{mX}$ with the theoretical predictions in Fig. \ref{fig.sketch}(c). The predicted $g$-factor for nearly free intra-layer MoSe$_2$ excitons is $g_A \approx -4$. A 22\% lower value for the moiré exciton yields $g_{mX} \approx -3.12$ which, according to Fig. \ref{fig.sketch}(c), requires an exciton momentum $K \approx 0.11$ \AA\,$^{-1}$. This corresponds to a confined state with energy roughly $\hbar^2 K^2 /2M \approx 40$ meV below that of the free exciton, where we used the exciton effective mass $M = 1.094 m_0$, as obtained from DFT. Indeed, this is in excellent agreement with the experimentally observed 43 meV separation between X$_A$ and mX exciton peaks in the PL spectra. Moreover, this energy is also in the same order of magnitude as the one observed for similar moiré-confined excitons in Ref. [\onlinecite{tran2019evidence}].  

The degree of valley polarization, defined as $P = (I^{\sigma_+}-I^{\sigma_-})/(I^{\sigma_+}+I^{\sigma_-})$, is shown in Fig. \ref{fig.gfactorsexp}(f-i) as a function of the field for each excitonic peak under circular polarization excitation. While $P$ is zero for all peaks in the absence of an applied magnetic field, the mX excitons exhibit a circular polarization as large as  $\approx$ 70\%  at field of 20 T, a nearly twice stronger polarization compared to the free X$_A$ exciton at this field. This corroborates the interpretation of X$_A$ as a predominantly intra-layer, nearly free, exciton, whose fast decay and relatively short lifetime limit the value of the field-induced valley polarization. On the other hand, the trion (X$^*$) has higher polarization degree, as it usually presents higher PL decay time and much longer intervalley scattering time \cite{Zhang2021-Polarization}. As the polarization degree depends on the ratio of the PL decay and intervalley scattering times, the high polarization degree for mX suggests an important increase of intervalley scattering times for the moiré confined exciton, as compared to that for X$_A$. However, further studies would be necessary to understand in more detail the observed magnetic field dependence of the polarization degree of the moiré confined exciton.




Notice that, despite the qualitative agreement between the $g$-factors of the different mX and X$_A$ peaks observed in our $\mu$PL experiment and their theoretically predicted values, the latter are quantitatively lower. In fact, $g$-factors of X$_A$ in MoSe$_2$ and WS$_2$ monolayers obtained by ab initio calculations are already slightly underestimated as compared to existing experimental data in the literature \cite{wozniak2020exciton}. Nevertheless, our model shows that the distinctly lower $g$-factors for mX states, consistently observed in the experiments, are a direct consequence of the fact that the exciton wave function is quantum confined by the moiré pattern. 

Assuming a possible inter-layer hybridization for mX \cite{alexeev2019resonantly}, where electrons are shared between both layers due to the low conduction band offset of this vdWHS, our DFT calculations predict an even lower exciton $g$-factor, as a consequence of the lower angular momentum for conduction band electrons in the WS$_2$ layer \cite{SM}. This indicates that the $g$-factor of these excitons could be further controlled e.g. by an external electric field, perpendicular to the vdWHS plane, which would push electrons and holes towards different layers. The fact that the exciton $g$-factors observed here in MoSe$_2$/WS$_2$ vdWHS are similar or even reduced as compared to those of excitons in monolayer MoSe$_2$, is also in stark contrast to the $g$-factors of excitons in other similar heterostructures, such as MoSe$_2$/WSe$_2$ vdWHS \cite{seyler2019signatures}, where exciton $g$-factors in the vdWHS are rather \textit{enhanced} as compared to those of either monolayer MoSe$_2$ or WSe$_2$.

For twist angles around 60$^{\circ}$, the inter-layer exciton $g$-factors are expected to be higher due to the sign change of the angular momentum for the conduction band states for these excitons \cite{wozniak2020exciton}. Indeed, significantly higher values for iX $g$-factors in MoSe$_2$/WSe$_2$ vdWHS with 60$^\circ$ twist angle are observed in Ref. \onlinecite{seyler2019signatures} for this reason. However, since the mX and X$_A$ states in MoSe$_2$/WS$_2$ vdWHS have a strong intra-layer component and low inter-layer contribution \cite{tang2021tuning}, our model estimates the $g$-factors in $\theta \approx 60^{\circ}$ twisted MoSe$_2$/WS$_2$ heterostructures to be similar to those for $\theta \approx 0^{\circ}$, in contrast to MoSe$_2$/WSe$_2$ vdWHS. Indeed, we experimentally verified this prediction with a 60$^\circ$ MoSe$_2$/WSe$_2$ vdWHS sample, for which the experimental results are detailed and discussed in the supplemental material. 

In conclusion, we have measured the polarization-resolved photoluminescence of exciton states in high quality hBN-encapsulated MoSe$_2$/WS$_2$ heterostructures with near-zero stacking angle, under perpendicular magnetic fields up to 20 T. Two PL peaks are identified as due to exciton states predominantly with an intra-layer MoSe$_2$ component: one originates from a free exciton state, with higher energy, whereas the lower energy peak is identified as the moiré-confined exciton. The energy separation between these peaks suggests a moiré confinement energy of 43 meV, with respect to the free exciton state, which matches theoretical estimates. Both exciton energies shift linearly with magnetic field, but the moiré-confined exciton exhibits reduced $g$-factors compared to the unconfined one. This is a consequence of the quantized center-of-mass momentum of the moiré-confined exciton, which makes it probe regions of the Brillouin zone where the angular momentum state of the electron-hole pair is significantly reduced, as demonstrated by our calculations. By generalization of this finding we envision different possible behavior of exciton $g-$factors at different stacking angles, or in vdWHS composed of other combinations of 2D TMDs, where exciton confinement can be further enhanced e.g. by application of an interlayer electrostatic bias \cite{yu2017moire}. As a common denominator, we postulate that moiré-confined excitons in vdW heterostructures are discernible through their $g$-factors, which offers a fresh perspective on further studies of moiré excitons, especially in twisted bilayers, or under controlled strain and gating, where further rich manifestations of exciton physics are expected. 

\textit{Acknowledgements} This work was supported by HFML-RU/NWO-I, member of the European Magnetic Field Laboratory (EMFL). This work was also financially supported by the Brazilian Council for Research (CNPq), through the PRONEX/FUNCAP, Universal, and PQ programs (311678/2020-3), and the Research Foundation - Flanders. CSB acknowledges the financial support from CAPES fellowship and YGG  acknowledges the financial support from FAPESP  (grants 2018/01808-5, 2019/14017-9 and 2019/23488-5) and CNPQ ( grants 311678/2020-3 and 426634/2018-7). TW acknowledges financial support from National Science Centre, Poland under grant 2021/41/N/ST3/04516. DFT calculations were performed with the support of the Interdisciplinary Centre for Mathematical and Computational Modelling (ICM) University of Warsaw and Center for Information Services and High Performance Computing (ZIH) at TU Dresden.

\bibliographystyle{apsrev4-2}
\bibliography{references}

\begin{thebibliography}{42}%
\makeatletter
\providecommand \@ifxundefined [1]{%
 \@ifx{#1\undefined}
}%
\providecommand \@ifnum [1]{%
 \ifnum #1\expandafter \@firstoftwo
 \else \expandafter \@secondoftwo
 \fi
}%
\providecommand \@ifx [1]{%
 \ifx #1\expandafter \@firstoftwo
 \else \expandafter \@secondoftwo
 \fi
}%
\providecommand \natexlab [1]{#1}%
\providecommand \enquote  [1]{``#1''}%
\providecommand \bibnamefont  [1]{#1}%
\providecommand \bibfnamefont [1]{#1}%
\providecommand \citenamefont [1]{#1}%
\providecommand \href@noop [0]{\@secondoftwo}%
\providecommand \href [0]{\begingroup \@sanitize@url \@href}%
\providecommand \@href[1]{\@@startlink{#1}\@@href}%
\providecommand \@@href[1]{\endgroup#1\@@endlink}%
\providecommand \@sanitize@url [0]{\catcode `\\12\catcode `\$12\catcode
  `\&12\catcode `\#12\catcode `\^12\catcode `\_12\catcode `\%12\relax}%
\providecommand \@@startlink[1]{}%
\providecommand \@@endlink[0]{}%
\providecommand \url  [0]{\begingroup\@sanitize@url \@url }%
\providecommand \@url [1]{\endgroup\@href {#1}{\urlprefix }}%
\providecommand \urlprefix  [0]{URL }%
\providecommand \Eprint [0]{\href }%
\providecommand \doibase [0]{https://doi.org/}%
\providecommand \selectlanguage [0]{\@gobble}%
\providecommand \bibinfo  [0]{\@secondoftwo}%
\providecommand \bibfield  [0]{\@secondoftwo}%
\providecommand \translation [1]{[#1]}%
\providecommand \BibitemOpen [0]{}%
\providecommand \bibitemStop [0]{}%
\providecommand \bibitemNoStop [0]{.\EOS\space}%
\providecommand \EOS [0]{\spacefactor3000\relax}%
\providecommand \BibitemShut  [1]{\csname bibitem#1\endcsname}%
\let\auto@bib@innerbib\@empty
\bibitem [{\citenamefont {Geim}\ and\ \citenamefont
  {Grigorieva}(2013)}]{geim2013van}%
  \BibitemOpen
  \bibfield  {author} {\bibinfo {author} {\bibfnamefont {A.~K.}\ \bibnamefont
  {Geim}}\ and\ \bibinfo {author} {\bibfnamefont {I.~V.}\ \bibnamefont
  {Grigorieva}},\ }\href@noop {} {\bibfield  {journal} {\bibinfo  {journal}
  {Nature}\ }\textbf {\bibinfo {volume} {499}},\ \bibinfo {pages} {419}
  (\bibinfo {year} {2013})}\BibitemShut {NoStop}%
\bibitem [{\citenamefont {Novoselov}\ \emph {et~al.}(2016)\citenamefont
  {Novoselov}, \citenamefont {Mishchenko}, \citenamefont {Carvalho},\ and\
  \citenamefont {Neto}}]{novoselov20162d}%
  \BibitemOpen
  \bibfield  {author} {\bibinfo {author} {\bibfnamefont {K.}~\bibnamefont
  {Novoselov}}, \bibinfo {author} {\bibfnamefont {A.}~\bibnamefont
  {Mishchenko}}, \bibinfo {author} {\bibfnamefont {A.}~\bibnamefont
  {Carvalho}},\ and\ \bibinfo {author} {\bibfnamefont {A.~C.}\ \bibnamefont
  {Neto}},\ }\href@noop {} {\bibfield  {journal} {\bibinfo  {journal}
  {Science}\ }\textbf {\bibinfo {volume} {353}} (\bibinfo {year}
  {2016})}\BibitemShut {NoStop}%
\bibitem [{\citenamefont {Mak}\ \emph {et~al.}(2010)\citenamefont {Mak},
  \citenamefont {Lee}, \citenamefont {Hone}, \citenamefont {Shan},\ and\
  \citenamefont {Heinz}}]{Mak2010}%
  \BibitemOpen
  \bibfield  {author} {\bibinfo {author} {\bibfnamefont {K.~F.}\ \bibnamefont
  {Mak}}, \bibinfo {author} {\bibfnamefont {C.}~\bibnamefont {Lee}}, \bibinfo
  {author} {\bibfnamefont {J.}~\bibnamefont {Hone}}, \bibinfo {author}
  {\bibfnamefont {J.}~\bibnamefont {Shan}},\ and\ \bibinfo {author}
  {\bibfnamefont {T.~F.}\ \bibnamefont {Heinz}},\ }\href
  {https://doi.org/10.1103/PhysRevLett.105.136805} {\bibfield  {journal}
  {\bibinfo  {journal} {Physical Review Letters}\ }\textbf {\bibinfo {volume}
  {105}},\ \bibinfo {pages} {2} (\bibinfo {year} {2010})}\BibitemShut {NoStop}%
\bibitem [{\citenamefont {Splendiani}\ \emph {et~al.}(2010)\citenamefont
  {Splendiani}, \citenamefont {Sun}, \citenamefont {Zhang}, \citenamefont {Li},
  \citenamefont {Kim}, \citenamefont {Chim}, \citenamefont {Galli},\ and\
  \citenamefont {Wang}}]{Splendiani2010}%
  \BibitemOpen
  \bibfield  {author} {\bibinfo {author} {\bibfnamefont {A.}~\bibnamefont
  {Splendiani}}, \bibinfo {author} {\bibfnamefont {L.}~\bibnamefont {Sun}},
  \bibinfo {author} {\bibfnamefont {Y.}~\bibnamefont {Zhang}}, \bibinfo
  {author} {\bibfnamefont {T.}~\bibnamefont {Li}}, \bibinfo {author}
  {\bibfnamefont {J.}~\bibnamefont {Kim}}, \bibinfo {author} {\bibfnamefont
  {C.~Y.}\ \bibnamefont {Chim}}, \bibinfo {author} {\bibfnamefont
  {G.}~\bibnamefont {Galli}},\ and\ \bibinfo {author} {\bibfnamefont
  {F.}~\bibnamefont {Wang}},\ }\href {https://doi.org/10.1021/nl903868w}
  {\bibfield  {journal} {\bibinfo  {journal} {Nano Letters}\ }\textbf {\bibinfo
  {volume} {10}},\ \bibinfo {pages} {1271} (\bibinfo {year}
  {2010})}\BibitemShut {NoStop}%
\bibitem [{\citenamefont {Xu}\ \emph {et~al.}(2014)\citenamefont {Xu},
  \citenamefont {Yao}, \citenamefont {Xiao},\ and\ \citenamefont
  {Heinz}}]{Xu2014}%
  \BibitemOpen
  \bibfield  {author} {\bibinfo {author} {\bibfnamefont {X.}~\bibnamefont
  {Xu}}, \bibinfo {author} {\bibfnamefont {W.}~\bibnamefont {Yao}}, \bibinfo
  {author} {\bibfnamefont {D.}~\bibnamefont {Xiao}},\ and\ \bibinfo {author}
  {\bibfnamefont {T.~F.}\ \bibnamefont {Heinz}},\ }\href
  {https://doi.org/10.1038/nphys2942} {\bibfield  {journal} {\bibinfo
  {journal} {Nature Physics}\ }\textbf {\bibinfo {volume} {10}},\ \bibinfo
  {pages} {343} (\bibinfo {year} {2014})}\BibitemShut {NoStop}%
\bibitem [{\citenamefont {Xiao}\ \emph {et~al.}(2012)\citenamefont {Xiao},
  \citenamefont {Liu}, \citenamefont {Feng}, \citenamefont {Xu},\ and\
  \citenamefont {Yao}}]{Xiao2012}%
  \BibitemOpen
  \bibfield  {author} {\bibinfo {author} {\bibfnamefont {D.}~\bibnamefont
  {Xiao}}, \bibinfo {author} {\bibfnamefont {G.~B.}\ \bibnamefont {Liu}},
  \bibinfo {author} {\bibfnamefont {W.}~\bibnamefont {Feng}}, \bibinfo {author}
  {\bibfnamefont {X.}~\bibnamefont {Xu}},\ and\ \bibinfo {author}
  {\bibfnamefont {W.}~\bibnamefont {Yao}},\ }\href
  {https://doi.org/10.1103/PhysRevLett.108.196802} {\bibfield  {journal}
  {\bibinfo  {journal} {Physical Review Letters}\ }\textbf {\bibinfo {volume}
  {108}},\ \bibinfo {pages} {1} (\bibinfo {year} {2012})}\BibitemShut {NoStop}%
\bibitem [{\citenamefont {Mak}\ \emph {et~al.}(2012)\citenamefont {Mak},
  \citenamefont {He}, \citenamefont {Shan},\ and\ \citenamefont
  {Heinz}}]{Mak2012}%
  \BibitemOpen
  \bibfield  {author} {\bibinfo {author} {\bibfnamefont {K.~F.}\ \bibnamefont
  {Mak}}, \bibinfo {author} {\bibfnamefont {K.}~\bibnamefont {He}}, \bibinfo
  {author} {\bibfnamefont {J.}~\bibnamefont {Shan}},\ and\ \bibinfo {author}
  {\bibfnamefont {T.~F.}\ \bibnamefont {Heinz}},\ }\href
  {https://doi.org/10.1038/nnano.2012.96} {\bibfield  {journal} {\bibinfo
  {journal} {Nature Nanotechnology}\ }\textbf {\bibinfo {volume} {7}},\
  \bibinfo {pages} {494} (\bibinfo {year} {2012})}\BibitemShut {NoStop}%
\bibitem [{\citenamefont {Zeng}\ \emph {et~al.}(2012)\citenamefont {Zeng},
  \citenamefont {Dai}, \citenamefont {Yao}, \citenamefont {Xiao},\ and\
  \citenamefont {Cui}}]{Zeng2012}%
  \BibitemOpen
  \bibfield  {author} {\bibinfo {author} {\bibfnamefont {H.}~\bibnamefont
  {Zeng}}, \bibinfo {author} {\bibfnamefont {J.}~\bibnamefont {Dai}}, \bibinfo
  {author} {\bibfnamefont {W.}~\bibnamefont {Yao}}, \bibinfo {author}
  {\bibfnamefont {D.}~\bibnamefont {Xiao}},\ and\ \bibinfo {author}
  {\bibfnamefont {X.}~\bibnamefont {Cui}},\ }\href
  {https://doi.org/10.1038/nnano.2012.95} {\bibfield  {journal} {\bibinfo
  {journal} {Nature Nanotechnology}\ }\textbf {\bibinfo {volume} {7}},\
  \bibinfo {pages} {490} (\bibinfo {year} {2012})}\BibitemShut {NoStop}%
\bibitem [{\citenamefont {Schaibley}\ \emph {et~al.}(2016)\citenamefont
  {Schaibley}, \citenamefont {Yu}, \citenamefont {Clark}, \citenamefont
  {Rivera}, \citenamefont {Ross}, \citenamefont {Seyler}, \citenamefont {Yao},\
  and\ \citenamefont {Xu}}]{Schaibley2016}%
  \BibitemOpen
  \bibfield  {author} {\bibinfo {author} {\bibfnamefont {J.~R.}\ \bibnamefont
  {Schaibley}}, \bibinfo {author} {\bibfnamefont {H.}~\bibnamefont {Yu}},
  \bibinfo {author} {\bibfnamefont {G.}~\bibnamefont {Clark}}, \bibinfo
  {author} {\bibfnamefont {P.}~\bibnamefont {Rivera}}, \bibinfo {author}
  {\bibfnamefont {J.~S.}\ \bibnamefont {Ross}}, \bibinfo {author}
  {\bibfnamefont {K.~L.}\ \bibnamefont {Seyler}}, \bibinfo {author}
  {\bibfnamefont {W.}~\bibnamefont {Yao}},\ and\ \bibinfo {author}
  {\bibfnamefont {X.}~\bibnamefont {Xu}},\ }\href
  {https://doi.org/10.1038/natrevmats.2016.55} {\bibfield  {journal} {\bibinfo
  {journal} {Nature Reviews Materials}\ }\textbf {\bibinfo {volume} {1}},\
  \bibinfo {pages} {1} (\bibinfo {year} {2016})}\BibitemShut {NoStop}%
\bibitem [{\citenamefont {Ishii}\ \emph {et~al.}(2019)\citenamefont {Ishii},
  \citenamefont {Yokoshi},\ and\ \citenamefont {Ishihara}}]{ishii2019optical}%
  \BibitemOpen
  \bibfield  {author} {\bibinfo {author} {\bibfnamefont {S.}~\bibnamefont
  {Ishii}}, \bibinfo {author} {\bibfnamefont {N.}~\bibnamefont {Yokoshi}},\
  and\ \bibinfo {author} {\bibfnamefont {H.}~\bibnamefont {Ishihara}},\ }in\
  \href@noop {} {\emph {\bibinfo {booktitle} {Journal of Physics: Conference
  Series}}},\ Vol.\ \bibinfo {volume} {1220}\ (\bibinfo {organization} {IOP
  Publishing},\ \bibinfo {year} {2019})\ p.\ \bibinfo {pages}
  {012056}\BibitemShut {NoStop}%
\bibitem [{\citenamefont {Aivazian}\ \emph {et~al.}(2015)\citenamefont
  {Aivazian}, \citenamefont {Gong}, \citenamefont {Jones}, \citenamefont {Chu},
  \citenamefont {Yan}, \citenamefont {Mandrus}, \citenamefont {Zhang},
  \citenamefont {Cobden}, \citenamefont {Yao},\ and\ \citenamefont
  {Xu}}]{Aivazian2015}%
  \BibitemOpen
  \bibfield  {author} {\bibinfo {author} {\bibfnamefont {G.}~\bibnamefont
  {Aivazian}}, \bibinfo {author} {\bibfnamefont {Z.}~\bibnamefont {Gong}},
  \bibinfo {author} {\bibfnamefont {A.~M.}\ \bibnamefont {Jones}}, \bibinfo
  {author} {\bibfnamefont {R.~L.}\ \bibnamefont {Chu}}, \bibinfo {author}
  {\bibfnamefont {J.}~\bibnamefont {Yan}}, \bibinfo {author} {\bibfnamefont
  {D.~G.}\ \bibnamefont {Mandrus}}, \bibinfo {author} {\bibfnamefont
  {C.}~\bibnamefont {Zhang}}, \bibinfo {author} {\bibfnamefont
  {D.}~\bibnamefont {Cobden}}, \bibinfo {author} {\bibfnamefont
  {W.}~\bibnamefont {Yao}},\ and\ \bibinfo {author} {\bibfnamefont
  {X.}~\bibnamefont {Xu}},\ }\href {https://doi.org/10.1038/nphys3201}
  {\bibfield  {journal} {\bibinfo  {journal} {Nature Physics}\ }\textbf
  {\bibinfo {volume} {11}},\ \bibinfo {pages} {148} (\bibinfo {year}
  {2015})}\BibitemShut {NoStop}%
\bibitem [{\citenamefont {Li}\ \emph {et~al.}(2014)\citenamefont {Li},
  \citenamefont {Ludwig}, \citenamefont {Low}, \citenamefont {Chernikov},
  \citenamefont {Cui}, \citenamefont {Arefe}, \citenamefont {Kim},
  \citenamefont {Zande}, \citenamefont {Rigosi}, \citenamefont {Hill},
  \citenamefont {Kim}, \citenamefont {Hone}, \citenamefont {Li}, \citenamefont
  {Smirnov},\ and\ \citenamefont {Heinz}}]{Li2014}%
  \BibitemOpen
  \bibfield  {author} {\bibinfo {author} {\bibfnamefont {Y.}~\bibnamefont
  {Li}}, \bibinfo {author} {\bibfnamefont {J.}~\bibnamefont {Ludwig}}, \bibinfo
  {author} {\bibfnamefont {T.}~\bibnamefont {Low}}, \bibinfo {author}
  {\bibfnamefont {A.}~\bibnamefont {Chernikov}}, \bibinfo {author}
  {\bibfnamefont {X.}~\bibnamefont {Cui}}, \bibinfo {author} {\bibfnamefont
  {G.}~\bibnamefont {Arefe}}, \bibinfo {author} {\bibfnamefont {Y.~D.}\
  \bibnamefont {Kim}}, \bibinfo {author} {\bibfnamefont {A.~M. V.~D.}\
  \bibnamefont {Zande}}, \bibinfo {author} {\bibfnamefont {A.}~\bibnamefont
  {Rigosi}}, \bibinfo {author} {\bibfnamefont {H.~M.}\ \bibnamefont {Hill}},
  \bibinfo {author} {\bibfnamefont {S.~H.}\ \bibnamefont {Kim}}, \bibinfo
  {author} {\bibfnamefont {J.}~\bibnamefont {Hone}}, \bibinfo {author}
  {\bibfnamefont {Z.}~\bibnamefont {Li}}, \bibinfo {author} {\bibfnamefont
  {D.}~\bibnamefont {Smirnov}},\ and\ \bibinfo {author} {\bibfnamefont {T.~F.}\
  \bibnamefont {Heinz}},\ }\href
  {https://doi.org/10.1103/PhysRevLett.113.266804} {\bibfield  {journal}
  {\bibinfo  {journal} {Physical Review Letters}\ }\textbf {\bibinfo {volume}
  {113}},\ \bibinfo {pages} {1} (\bibinfo {year} {2014})}\BibitemShut {NoStop}%
\bibitem [{\citenamefont {Srivastava}\ \emph {et~al.}(2015)\citenamefont
  {Srivastava}, \citenamefont {Sidler}, \citenamefont {Allain}, \citenamefont
  {Lembke}, \citenamefont {Kis},\ and\ \citenamefont
  {Imamoglu}}]{Srivastava2015}%
  \BibitemOpen
  \bibfield  {author} {\bibinfo {author} {\bibfnamefont {A.}~\bibnamefont
  {Srivastava}}, \bibinfo {author} {\bibfnamefont {M.}~\bibnamefont {Sidler}},
  \bibinfo {author} {\bibfnamefont {A.~V.}\ \bibnamefont {Allain}}, \bibinfo
  {author} {\bibfnamefont {D.~S.}\ \bibnamefont {Lembke}}, \bibinfo {author}
  {\bibfnamefont {A.}~\bibnamefont {Kis}},\ and\ \bibinfo {author}
  {\bibfnamefont {A.}~\bibnamefont {Imamoglu}},\ }\href
  {https://doi.org/10.1038/nphys3203} {\bibfield  {journal} {\bibinfo
  {journal} {Nature Physics}\ }\textbf {\bibinfo {volume} {11}},\ \bibinfo
  {pages} {141} (\bibinfo {year} {2015})}\BibitemShut {NoStop}%
\bibitem [{\citenamefont {Macneill}\ \emph {et~al.}(2015)\citenamefont
  {Macneill}, \citenamefont {Heikes}, \citenamefont {Mak}, \citenamefont
  {Anderson}, \citenamefont {Kormányos}, \citenamefont {Zólyomi},
  \citenamefont {Park},\ and\ \citenamefont {Ralph}}]{Macneill2015}%
  \BibitemOpen
  \bibfield  {author} {\bibinfo {author} {\bibfnamefont {D.}~\bibnamefont
  {Macneill}}, \bibinfo {author} {\bibfnamefont {C.}~\bibnamefont {Heikes}},
  \bibinfo {author} {\bibfnamefont {K.~F.}\ \bibnamefont {Mak}}, \bibinfo
  {author} {\bibfnamefont {Z.}~\bibnamefont {Anderson}}, \bibinfo {author}
  {\bibfnamefont {A.}~\bibnamefont {Kormányos}}, \bibinfo {author}
  {\bibfnamefont {V.}~\bibnamefont {Zólyomi}}, \bibinfo {author}
  {\bibfnamefont {J.}~\bibnamefont {Park}},\ and\ \bibinfo {author}
  {\bibfnamefont {D.~C.}\ \bibnamefont {Ralph}},\ }\href
  {https://doi.org/10.1103/PhysRevLett.114.037401} {\bibfield  {journal}
  {\bibinfo  {journal} {Physical Review Letters}\ }\textbf {\bibinfo {volume}
  {114}},\ \bibinfo {pages} {1} (\bibinfo {year} {2015})}\BibitemShut {NoStop}%
\bibitem [{\citenamefont {Wang}\ \emph {et~al.}(2015)\citenamefont {Wang},
  \citenamefont {Bouet}, \citenamefont {Glazov}, \citenamefont {Amand},
  \citenamefont {Ivchenko}, \citenamefont {Palleau}, \citenamefont {Marie},\
  and\ \citenamefont {Urbaszek}}]{Wang2015}%
  \BibitemOpen
  \bibfield  {author} {\bibinfo {author} {\bibfnamefont {G.}~\bibnamefont
  {Wang}}, \bibinfo {author} {\bibfnamefont {L.}~\bibnamefont {Bouet}},
  \bibinfo {author} {\bibfnamefont {M.~M.}\ \bibnamefont {Glazov}}, \bibinfo
  {author} {\bibfnamefont {T.}~\bibnamefont {Amand}}, \bibinfo {author}
  {\bibfnamefont {E.~L.}\ \bibnamefont {Ivchenko}}, \bibinfo {author}
  {\bibfnamefont {E.}~\bibnamefont {Palleau}}, \bibinfo {author} {\bibfnamefont
  {X.}~\bibnamefont {Marie}},\ and\ \bibinfo {author} {\bibfnamefont
  {B.}~\bibnamefont {Urbaszek}},\ }\href
  {https://doi.org/10.1088/2053-1583/2/3/034002} {\bibfield  {journal}
  {\bibinfo  {journal} {2D Materials}\ }\textbf {\bibinfo {volume} {2}},\
  \bibinfo {pages} {34002} (\bibinfo {year} {2015})}\BibitemShut {NoStop}%
\bibitem [{\citenamefont {Mitioglu}\ \emph {et~al.}(2015)\citenamefont
  {Mitioglu}, \citenamefont {Plochocka}, \citenamefont {Aguila}, \citenamefont
  {Christianen}, \citenamefont {Deligeorgis}, \citenamefont {Anghel},
  \citenamefont {Kulyuk},\ and\ \citenamefont {Maude}}]{Mitioglu2015}%
  \BibitemOpen
  \bibfield  {author} {\bibinfo {author} {\bibfnamefont {A.~A.}\ \bibnamefont
  {Mitioglu}}, \bibinfo {author} {\bibfnamefont {P.}~\bibnamefont {Plochocka}},
  \bibinfo {author} {\bibfnamefont {G.~D.}\ \bibnamefont {Aguila}}, \bibinfo
  {author} {\bibfnamefont {P.~C.}\ \bibnamefont {Christianen}}, \bibinfo
  {author} {\bibfnamefont {G.}~\bibnamefont {Deligeorgis}}, \bibinfo {author}
  {\bibfnamefont {S.}~\bibnamefont {Anghel}}, \bibinfo {author} {\bibfnamefont
  {L.}~\bibnamefont {Kulyuk}},\ and\ \bibinfo {author} {\bibfnamefont {D.~K.}\
  \bibnamefont {Maude}},\ }\href {https://doi.org/10.1021/acs.nanolett.5b00626}
  {\bibfield  {journal} {\bibinfo  {journal} {Nano Letters}\ }\textbf {\bibinfo
  {volume} {15}},\ \bibinfo {pages} {4387} (\bibinfo {year}
  {2015})}\BibitemShut {NoStop}%
\bibitem [{\citenamefont {Stier}\ \emph {et~al.}(2016)\citenamefont {Stier},
  \citenamefont {McCreary}, \citenamefont {Jonker}, \citenamefont {Kono},\ and\
  \citenamefont {Crooker}}]{Stier2016}%
  \BibitemOpen
  \bibfield  {author} {\bibinfo {author} {\bibfnamefont {A.~V.}\ \bibnamefont
  {Stier}}, \bibinfo {author} {\bibfnamefont {K.~M.}\ \bibnamefont {McCreary}},
  \bibinfo {author} {\bibfnamefont {B.~T.}\ \bibnamefont {Jonker}}, \bibinfo
  {author} {\bibfnamefont {J.}~\bibnamefont {Kono}},\ and\ \bibinfo {author}
  {\bibfnamefont {S.~A.}\ \bibnamefont {Crooker}},\ }\href
  {https://doi.org/10.1038/ncomms10643} {\bibfield  {journal} {\bibinfo
  {journal} {Nature Communications}\ }\textbf {\bibinfo {volume} {7}},\
  \bibinfo {pages} {1} (\bibinfo {year} {2016})}\BibitemShut {NoStop}%
\bibitem [{\citenamefont {Plechinger}\ \emph {et~al.}(2016)\citenamefont
  {Plechinger}, \citenamefont {Nagler}, \citenamefont {Arora}, \citenamefont
  {Águila}, \citenamefont {Ballottin}, \citenamefont {Frank}, \citenamefont
  {Steinleitner}, \citenamefont {Gmitra}, \citenamefont {Fabian}, \citenamefont
  {Christianen}, \citenamefont {Bratschitsch}, \citenamefont {Schüller},\ and\
  \citenamefont {Korn}}]{Plechinger2016}%
  \BibitemOpen
  \bibfield  {author} {\bibinfo {author} {\bibfnamefont {G.}~\bibnamefont
  {Plechinger}}, \bibinfo {author} {\bibfnamefont {P.}~\bibnamefont {Nagler}},
  \bibinfo {author} {\bibfnamefont {A.}~\bibnamefont {Arora}}, \bibinfo
  {author} {\bibfnamefont {A.~G.~D.}\ \bibnamefont {Águila}}, \bibinfo
  {author} {\bibfnamefont {M.~V.}\ \bibnamefont {Ballottin}}, \bibinfo {author}
  {\bibfnamefont {T.}~\bibnamefont {Frank}}, \bibinfo {author} {\bibfnamefont
  {P.}~\bibnamefont {Steinleitner}}, \bibinfo {author} {\bibfnamefont
  {M.}~\bibnamefont {Gmitra}}, \bibinfo {author} {\bibfnamefont
  {J.}~\bibnamefont {Fabian}}, \bibinfo {author} {\bibfnamefont {P.~C.}\
  \bibnamefont {Christianen}}, \bibinfo {author} {\bibfnamefont
  {R.}~\bibnamefont {Bratschitsch}}, \bibinfo {author} {\bibfnamefont
  {C.}~\bibnamefont {Schüller}},\ and\ \bibinfo {author} {\bibfnamefont
  {T.}~\bibnamefont {Korn}},\ }\href
  {https://doi.org/10.1021/acs.nanolett.6b04171} {\bibfield  {journal}
  {\bibinfo  {journal} {Nano Letters}\ }\textbf {\bibinfo {volume} {16}},\
  \bibinfo {pages} {7899} (\bibinfo {year} {2016})}\BibitemShut {NoStop}%
\bibitem [{\citenamefont {Rivera}\ \emph {et~al.}(2016)\citenamefont {Rivera},
  \citenamefont {Seyler}, \citenamefont {Yu}, \citenamefont {Schaibley},
  \citenamefont {Yan}, \citenamefont {Mandrus}, \citenamefont {Yao},\ and\
  \citenamefont {Xu}}]{Rivera2016}%
  \BibitemOpen
  \bibfield  {author} {\bibinfo {author} {\bibfnamefont {P.}~\bibnamefont
  {Rivera}}, \bibinfo {author} {\bibfnamefont {K.~L.}\ \bibnamefont {Seyler}},
  \bibinfo {author} {\bibfnamefont {H.}~\bibnamefont {Yu}}, \bibinfo {author}
  {\bibfnamefont {J.~R.}\ \bibnamefont {Schaibley}}, \bibinfo {author}
  {\bibfnamefont {J.}~\bibnamefont {Yan}}, \bibinfo {author} {\bibfnamefont
  {D.~G.}\ \bibnamefont {Mandrus}}, \bibinfo {author} {\bibfnamefont
  {W.}~\bibnamefont {Yao}},\ and\ \bibinfo {author} {\bibfnamefont
  {X.}~\bibnamefont {Xu}},\ }\href {https://doi.org/10.1126/science.aac7820}
  {\bibfield  {journal} {\bibinfo  {journal} {Science}\ }\textbf {\bibinfo
  {volume} {351}},\ \bibinfo {pages} {688} (\bibinfo {year}
  {2016})}\BibitemShut {NoStop}%
\bibitem [{\citenamefont {Nayak}\ \emph {et~al.}(2017)\citenamefont {Nayak},
  \citenamefont {Horbatenko}, \citenamefont {Ahn}, \citenamefont {Kim},
  \citenamefont {Lee}, \citenamefont {Ma}, \citenamefont {Jang}, \citenamefont
  {Lim}, \citenamefont {Kim}, \citenamefont {Ryu}, \citenamefont {Cheong},
  \citenamefont {Park},\ and\ \citenamefont {Shin}}]{Nayak2017}%
  \BibitemOpen
  \bibfield  {author} {\bibinfo {author} {\bibfnamefont {P.~K.}\ \bibnamefont
  {Nayak}}, \bibinfo {author} {\bibfnamefont {Y.}~\bibnamefont {Horbatenko}},
  \bibinfo {author} {\bibfnamefont {S.}~\bibnamefont {Ahn}}, \bibinfo {author}
  {\bibfnamefont {G.}~\bibnamefont {Kim}}, \bibinfo {author} {\bibfnamefont
  {J.~U.}\ \bibnamefont {Lee}}, \bibinfo {author} {\bibfnamefont {K.~Y.}\
  \bibnamefont {Ma}}, \bibinfo {author} {\bibfnamefont {A.~R.}\ \bibnamefont
  {Jang}}, \bibinfo {author} {\bibfnamefont {H.}~\bibnamefont {Lim}}, \bibinfo
  {author} {\bibfnamefont {D.}~\bibnamefont {Kim}}, \bibinfo {author}
  {\bibfnamefont {S.}~\bibnamefont {Ryu}}, \bibinfo {author} {\bibfnamefont
  {H.}~\bibnamefont {Cheong}}, \bibinfo {author} {\bibfnamefont
  {N.}~\bibnamefont {Park}},\ and\ \bibinfo {author} {\bibfnamefont {H.~S.}\
  \bibnamefont {Shin}},\ }\href {https://doi.org/10.1021/acsnano.7b00640}
  {\bibfield  {journal} {\bibinfo  {journal} {ACS Nano}\ }\textbf {\bibinfo
  {volume} {11}},\ \bibinfo {pages} {4041} (\bibinfo {year}
  {2017})}\BibitemShut {NoStop}%
\bibitem [{\citenamefont {Nagler}\ \emph {et~al.}(2017)\citenamefont {Nagler},
  \citenamefont {Ballottin}, \citenamefont {Mitioglu}, \citenamefont
  {Mooshammer}, \citenamefont {Paradiso}, \citenamefont {Strunk}, \citenamefont
  {Huber}, \citenamefont {Chernikov}, \citenamefont {Christianen},
  \citenamefont {Schüller},\ and\ \citenamefont {Korn}}]{Nagler2017}%
  \BibitemOpen
  \bibfield  {author} {\bibinfo {author} {\bibfnamefont {P.}~\bibnamefont
  {Nagler}}, \bibinfo {author} {\bibfnamefont {M.~V.}\ \bibnamefont
  {Ballottin}}, \bibinfo {author} {\bibfnamefont {A.~A.}\ \bibnamefont
  {Mitioglu}}, \bibinfo {author} {\bibfnamefont {F.}~\bibnamefont
  {Mooshammer}}, \bibinfo {author} {\bibfnamefont {N.}~\bibnamefont
  {Paradiso}}, \bibinfo {author} {\bibfnamefont {C.}~\bibnamefont {Strunk}},
  \bibinfo {author} {\bibfnamefont {R.}~\bibnamefont {Huber}}, \bibinfo
  {author} {\bibfnamefont {A.}~\bibnamefont {Chernikov}}, \bibinfo {author}
  {\bibfnamefont {P.~C.}\ \bibnamefont {Christianen}}, \bibinfo {author}
  {\bibfnamefont {C.}~\bibnamefont {Schüller}},\ and\ \bibinfo {author}
  {\bibfnamefont {T.}~\bibnamefont {Korn}},\ }\href
  {https://doi.org/10.1038/s41467-017-01748-1} {\bibfield  {journal} {\bibinfo
  {journal} {Nature Communications}\ }\textbf {\bibinfo {volume} {8}},\
  \bibinfo {pages} {1} (\bibinfo {year} {2017})}\BibitemShut {NoStop}%
\bibitem [{\citenamefont {Förg}\ \emph {et~al.}(2021)\citenamefont {Förg},
  \citenamefont {Baimuratov}, \citenamefont {Kruchinin}, \citenamefont {Vovk},
  \citenamefont {Scherzer}, \citenamefont {Förste}, \citenamefont {Funk},
  \citenamefont {Watanabe}, \citenamefont {Taniguchi},\ and\ \citenamefont
  {Högele}}]{Forg2021}%
  \BibitemOpen
  \bibfield  {author} {\bibinfo {author} {\bibfnamefont {M.}~\bibnamefont
  {Förg}}, \bibinfo {author} {\bibfnamefont {A.~S.}\ \bibnamefont
  {Baimuratov}}, \bibinfo {author} {\bibfnamefont {S.~Y.}\ \bibnamefont
  {Kruchinin}}, \bibinfo {author} {\bibfnamefont {I.~A.}\ \bibnamefont {Vovk}},
  \bibinfo {author} {\bibfnamefont {J.}~\bibnamefont {Scherzer}}, \bibinfo
  {author} {\bibfnamefont {J.}~\bibnamefont {Förste}}, \bibinfo {author}
  {\bibfnamefont {V.}~\bibnamefont {Funk}}, \bibinfo {author} {\bibfnamefont
  {K.}~\bibnamefont {Watanabe}}, \bibinfo {author} {\bibfnamefont
  {T.}~\bibnamefont {Taniguchi}},\ and\ \bibinfo {author} {\bibfnamefont
  {A.}~\bibnamefont {Högele}},\ }\href
  {https://doi.org/10.1038/s41467-021-21822-z} {\bibfield  {journal} {\bibinfo
  {journal} {Nature Communications}\ }\textbf {\bibinfo {volume} {12}},\
  \bibinfo {pages} {1} (\bibinfo {year} {2021})}\BibitemShut {NoStop}%
\bibitem [{\citenamefont {Rivera}\ \emph {et~al.}(2018)\citenamefont {Rivera},
  \citenamefont {Yu}, \citenamefont {Seyler}, \citenamefont {Wilson},
  \citenamefont {Yao},\ and\ \citenamefont {Xu}}]{Rivera2018}%
  \BibitemOpen
  \bibfield  {author} {\bibinfo {author} {\bibfnamefont {P.}~\bibnamefont
  {Rivera}}, \bibinfo {author} {\bibfnamefont {H.}~\bibnamefont {Yu}}, \bibinfo
  {author} {\bibfnamefont {K.~L.}\ \bibnamefont {Seyler}}, \bibinfo {author}
  {\bibfnamefont {N.~P.}\ \bibnamefont {Wilson}}, \bibinfo {author}
  {\bibfnamefont {W.}~\bibnamefont {Yao}},\ and\ \bibinfo {author}
  {\bibfnamefont {X.}~\bibnamefont {Xu}},\ }\href
  {https://doi.org/10.1038/s41565-018-0193-0} {\bibfield  {journal} {\bibinfo
  {journal} {Nature Nanotechnology}\ }\textbf {\bibinfo {volume} {13}},\
  \bibinfo {pages} {1004} (\bibinfo {year} {2018})}\BibitemShut {NoStop}%
\bibitem [{\citenamefont {Tartakovskii}(2020)}]{Tartakovskii2020}%
  \BibitemOpen
  \bibfield  {author} {\bibinfo {author} {\bibfnamefont {A.}~\bibnamefont
  {Tartakovskii}},\ }\href {https://doi.org/10.1038/s42254-019-0136-1}
  {\bibfield  {journal} {\bibinfo  {journal} {Nature Reviews Physics}\ }\textbf
  {\bibinfo {volume} {2}},\ \bibinfo {pages} {8} (\bibinfo {year}
  {2020})}\BibitemShut {NoStop}%
\bibitem [{\citenamefont {Kuwabara}\ \emph {et~al.}(1990)\citenamefont
  {Kuwabara}, \citenamefont {Clarke},\ and\ \citenamefont
  {Smith}}]{kuwabara1990anomalous}%
  \BibitemOpen
  \bibfield  {author} {\bibinfo {author} {\bibfnamefont {M.}~\bibnamefont
  {Kuwabara}}, \bibinfo {author} {\bibfnamefont {D.~R.}\ \bibnamefont
  {Clarke}},\ and\ \bibinfo {author} {\bibfnamefont {D.}~\bibnamefont
  {Smith}},\ }\href@noop {} {\bibfield  {journal} {\bibinfo  {journal} {Applied
  physics letters}\ }\textbf {\bibinfo {volume} {56}},\ \bibinfo {pages} {2396}
  (\bibinfo {year} {1990})}\BibitemShut {NoStop}%
\bibitem [{\citenamefont {Tran}\ \emph {et~al.}(2019)\citenamefont {Tran},
  \citenamefont {Moody}, \citenamefont {Wu}, \citenamefont {Lu}, \citenamefont
  {Choi}, \citenamefont {Kim}, \citenamefont {Rai}, \citenamefont {Sanchez},
  \citenamefont {Quan}, \citenamefont {Singh} \emph
  {et~al.}}]{tran2019evidence}%
  \BibitemOpen
  \bibfield  {author} {\bibinfo {author} {\bibfnamefont {K.}~\bibnamefont
  {Tran}}, \bibinfo {author} {\bibfnamefont {G.}~\bibnamefont {Moody}},
  \bibinfo {author} {\bibfnamefont {F.}~\bibnamefont {Wu}}, \bibinfo {author}
  {\bibfnamefont {X.}~\bibnamefont {Lu}}, \bibinfo {author} {\bibfnamefont
  {J.}~\bibnamefont {Choi}}, \bibinfo {author} {\bibfnamefont {K.}~\bibnamefont
  {Kim}}, \bibinfo {author} {\bibfnamefont {A.}~\bibnamefont {Rai}}, \bibinfo
  {author} {\bibfnamefont {D.~A.}\ \bibnamefont {Sanchez}}, \bibinfo {author}
  {\bibfnamefont {J.}~\bibnamefont {Quan}}, \bibinfo {author} {\bibfnamefont
  {A.}~\bibnamefont {Singh}}, \emph {et~al.},\ }\href@noop {} {\bibfield
  {journal} {\bibinfo  {journal} {Nature}\ }\textbf {\bibinfo {volume} {567}},\
  \bibinfo {pages} {71} (\bibinfo {year} {2019})}\BibitemShut {NoStop}%
\bibitem [{\citenamefont {Baek}\ \emph {et~al.}(2020)\citenamefont {Baek},
  \citenamefont {Brotons-Gisbert}, \citenamefont {Koong}, \citenamefont
  {Campbell}, \citenamefont {Rambach}, \citenamefont {Watanabe}, \citenamefont
  {Taniguchi},\ and\ \citenamefont {Gerardot}}]{baek2020highly}%
  \BibitemOpen
  \bibfield  {author} {\bibinfo {author} {\bibfnamefont {H.}~\bibnamefont
  {Baek}}, \bibinfo {author} {\bibfnamefont {M.}~\bibnamefont
  {Brotons-Gisbert}}, \bibinfo {author} {\bibfnamefont {Z.}~\bibnamefont
  {Koong}}, \bibinfo {author} {\bibfnamefont {A.}~\bibnamefont {Campbell}},
  \bibinfo {author} {\bibfnamefont {M.}~\bibnamefont {Rambach}}, \bibinfo
  {author} {\bibfnamefont {K.}~\bibnamefont {Watanabe}}, \bibinfo {author}
  {\bibfnamefont {T.}~\bibnamefont {Taniguchi}},\ and\ \bibinfo {author}
  {\bibfnamefont {B.~D.}\ \bibnamefont {Gerardot}},\ }\href@noop {} {\bibfield
  {journal} {\bibinfo  {journal} {Science advances}\ }\textbf {\bibinfo
  {volume} {6}},\ \bibinfo {pages} {eaba8526} (\bibinfo {year}
  {2020})}\BibitemShut {NoStop}%
\bibitem [{\citenamefont {Huang}\ \emph {et~al.}(2022)\citenamefont {Huang},
  \citenamefont {Choi}, \citenamefont {Shih},\ and\ \citenamefont
  {Li}}]{huang2022excitons}%
  \BibitemOpen
  \bibfield  {author} {\bibinfo {author} {\bibfnamefont {D.}~\bibnamefont
  {Huang}}, \bibinfo {author} {\bibfnamefont {J.}~\bibnamefont {Choi}},
  \bibinfo {author} {\bibfnamefont {C.-K.}\ \bibnamefont {Shih}},\ and\
  \bibinfo {author} {\bibfnamefont {X.}~\bibnamefont {Li}},\ }\href@noop {}
  {\bibfield  {journal} {\bibinfo  {journal} {Nature Nanotechnology}\ ,\
  \bibinfo {pages} {1}} (\bibinfo {year} {2022})}\BibitemShut {NoStop}%
\bibitem [{\citenamefont {Seyler}\ \emph {et~al.}(2019)\citenamefont {Seyler},
  \citenamefont {Rivera}, \citenamefont {Yu}, \citenamefont {Wilson},
  \citenamefont {Ray}, \citenamefont {Mandrus}, \citenamefont {Yan},
  \citenamefont {Yao},\ and\ \citenamefont {Xu}}]{seyler2019signatures}%
  \BibitemOpen
  \bibfield  {author} {\bibinfo {author} {\bibfnamefont {K.~L.}\ \bibnamefont
  {Seyler}}, \bibinfo {author} {\bibfnamefont {P.}~\bibnamefont {Rivera}},
  \bibinfo {author} {\bibfnamefont {H.}~\bibnamefont {Yu}}, \bibinfo {author}
  {\bibfnamefont {N.~P.}\ \bibnamefont {Wilson}}, \bibinfo {author}
  {\bibfnamefont {E.~L.}\ \bibnamefont {Ray}}, \bibinfo {author} {\bibfnamefont
  {D.~G.}\ \bibnamefont {Mandrus}}, \bibinfo {author} {\bibfnamefont
  {J.}~\bibnamefont {Yan}}, \bibinfo {author} {\bibfnamefont {W.}~\bibnamefont
  {Yao}},\ and\ \bibinfo {author} {\bibfnamefont {X.}~\bibnamefont {Xu}},\
  }\href@noop {} {\bibfield  {journal} {\bibinfo  {journal} {Nature}\ }\textbf
  {\bibinfo {volume} {567}},\ \bibinfo {pages} {66} (\bibinfo {year}
  {2019})}\BibitemShut {NoStop}%
\bibitem [{\citenamefont {F{\"o}rg}\ \emph {et~al.}(2021)\citenamefont
  {F{\"o}rg}, \citenamefont {Baimuratov}, \citenamefont {Kruchinin},
  \citenamefont {Vovk}, \citenamefont {Scherzer}, \citenamefont {F{\"o}rste},
  \citenamefont {Funk}, \citenamefont {Watanabe}, \citenamefont {Taniguchi},\
  and\ \citenamefont {H{\"o}gele}}]{forg2021moire}%
  \BibitemOpen
  \bibfield  {author} {\bibinfo {author} {\bibfnamefont {M.}~\bibnamefont
  {F{\"o}rg}}, \bibinfo {author} {\bibfnamefont {A.~S.}\ \bibnamefont
  {Baimuratov}}, \bibinfo {author} {\bibfnamefont {S.~Y.}\ \bibnamefont
  {Kruchinin}}, \bibinfo {author} {\bibfnamefont {I.~A.}\ \bibnamefont {Vovk}},
  \bibinfo {author} {\bibfnamefont {J.}~\bibnamefont {Scherzer}}, \bibinfo
  {author} {\bibfnamefont {J.}~\bibnamefont {F{\"o}rste}}, \bibinfo {author}
  {\bibfnamefont {V.}~\bibnamefont {Funk}}, \bibinfo {author} {\bibfnamefont
  {K.}~\bibnamefont {Watanabe}}, \bibinfo {author} {\bibfnamefont
  {T.}~\bibnamefont {Taniguchi}},\ and\ \bibinfo {author} {\bibfnamefont
  {A.}~\bibnamefont {H{\"o}gele}},\ }\href@noop {} {\bibfield  {journal}
  {\bibinfo  {journal} {Nature communications}\ }\textbf {\bibinfo {volume}
  {12}},\ \bibinfo {pages} {1} (\bibinfo {year} {2021})}\BibitemShut {NoStop}%
\bibitem [{\citenamefont {Wo{\'z}niak}\ \emph {et~al.}(2020)\citenamefont
  {Wo{\'z}niak}, \citenamefont {Junior}, \citenamefont {Seifert}, \citenamefont
  {Chaves},\ and\ \citenamefont {Kunstmann}}]{wozniak2020exciton}%
  \BibitemOpen
  \bibfield  {author} {\bibinfo {author} {\bibfnamefont {T.}~\bibnamefont
  {Wo{\'z}niak}}, \bibinfo {author} {\bibfnamefont {P.~E.~F.}\ \bibnamefont
  {Junior}}, \bibinfo {author} {\bibfnamefont {G.}~\bibnamefont {Seifert}},
  \bibinfo {author} {\bibfnamefont {A.}~\bibnamefont {Chaves}},\ and\ \bibinfo
  {author} {\bibfnamefont {J.}~\bibnamefont {Kunstmann}},\ }\href@noop {}
  {\bibfield  {journal} {\bibinfo  {journal} {Physical Review B}\ }\textbf
  {\bibinfo {volume} {101}},\ \bibinfo {pages} {235408} (\bibinfo {year}
  {2020})}\BibitemShut {NoStop}%
\bibitem [{\citenamefont {Deilmann}\ \emph {et~al.}(2020)\citenamefont
  {Deilmann}, \citenamefont {Kr{\"u}ger},\ and\ \citenamefont
  {Rohlfing}}]{deilmann2020ab}%
  \BibitemOpen
  \bibfield  {author} {\bibinfo {author} {\bibfnamefont {T.}~\bibnamefont
  {Deilmann}}, \bibinfo {author} {\bibfnamefont {P.}~\bibnamefont
  {Kr{\"u}ger}},\ and\ \bibinfo {author} {\bibfnamefont {M.}~\bibnamefont
  {Rohlfing}},\ }\href@noop {} {\bibfield  {journal} {\bibinfo  {journal}
  {Physical review letters}\ }\textbf {\bibinfo {volume} {124}},\ \bibinfo
  {pages} {226402} (\bibinfo {year} {2020})}\BibitemShut {NoStop}%
\bibitem [{\citenamefont {Chen}\ \emph {et~al.}(2019)\citenamefont {Chen},
  \citenamefont {Lu}, \citenamefont {Goldstein}, \citenamefont {Tong},
  \citenamefont {Chaves}, \citenamefont {Kunstmann}, \citenamefont
  {Cavalcante}, \citenamefont {Wozniak}, \citenamefont {Seifert}, \citenamefont
  {Reichman} \emph {et~al.}}]{chen2019luminescent}%
  \BibitemOpen
  \bibfield  {author} {\bibinfo {author} {\bibfnamefont {S.-Y.}\ \bibnamefont
  {Chen}}, \bibinfo {author} {\bibfnamefont {Z.}~\bibnamefont {Lu}}, \bibinfo
  {author} {\bibfnamefont {T.}~\bibnamefont {Goldstein}}, \bibinfo {author}
  {\bibfnamefont {J.}~\bibnamefont {Tong}}, \bibinfo {author} {\bibfnamefont
  {A.}~\bibnamefont {Chaves}}, \bibinfo {author} {\bibfnamefont
  {J.}~\bibnamefont {Kunstmann}}, \bibinfo {author} {\bibfnamefont
  {L.}~\bibnamefont {Cavalcante}}, \bibinfo {author} {\bibfnamefont
  {T.}~\bibnamefont {Wozniak}}, \bibinfo {author} {\bibfnamefont
  {G.}~\bibnamefont {Seifert}}, \bibinfo {author} {\bibfnamefont
  {D.}~\bibnamefont {Reichman}}, \emph {et~al.},\ }\href@noop {} {\bibfield
  {journal} {\bibinfo  {journal} {Nano letters}\ }\textbf {\bibinfo {volume}
  {19}},\ \bibinfo {pages} {2464} (\bibinfo {year} {2019})}\BibitemShut
  {NoStop}%
\bibitem [{SM()}]{SM}%
  \BibitemOpen
  \href@noop {} {\bibinfo  {journal} {See Supplemental Material for additional
  measurements at different positions in the sample}\ }\BibitemShut {NoStop}%
\bibitem [{\citenamefont {Alexeev}\ \emph {et~al.}(2019)\citenamefont
  {Alexeev}, \citenamefont {Ruiz-Tijerina}, \citenamefont {Danovich},
  \citenamefont {Hamer}, \citenamefont {Terry}, \citenamefont {Nayak},
  \citenamefont {Ahn}, \citenamefont {Pak}, \citenamefont {Lee}, \citenamefont
  {Sohn} \emph {et~al.}}]{alexeev2019resonantly}%
  \BibitemOpen
\bibfield  {journal} {  }\bibfield  {author} {\bibinfo {author} {\bibfnamefont
  {E.~M.}\ \bibnamefont {Alexeev}}, \bibinfo {author} {\bibfnamefont {D.~A.}\
  \bibnamefont {Ruiz-Tijerina}}, \bibinfo {author} {\bibfnamefont
  {M.}~\bibnamefont {Danovich}}, \bibinfo {author} {\bibfnamefont {M.~J.}\
  \bibnamefont {Hamer}}, \bibinfo {author} {\bibfnamefont {D.~J.}\ \bibnamefont
  {Terry}}, \bibinfo {author} {\bibfnamefont {P.~K.}\ \bibnamefont {Nayak}},
  \bibinfo {author} {\bibfnamefont {S.}~\bibnamefont {Ahn}}, \bibinfo {author}
  {\bibfnamefont {S.}~\bibnamefont {Pak}}, \bibinfo {author} {\bibfnamefont
  {J.}~\bibnamefont {Lee}}, \bibinfo {author} {\bibfnamefont {J.~I.}\
  \bibnamefont {Sohn}}, \emph {et~al.},\ }\href@noop {} {\bibfield  {journal}
  {\bibinfo  {journal} {Nature}\ }\textbf {\bibinfo {volume} {567}},\ \bibinfo
  {pages} {81} (\bibinfo {year} {2019})}\BibitemShut {NoStop}%
\bibitem [{\citenamefont {Zhang}\ \emph {et~al.}(2020)\citenamefont {Zhang},
  \citenamefont {Zhang}, \citenamefont {Wu}, \citenamefont {Wang},
  \citenamefont {Gogna}, \citenamefont {Hou}, \citenamefont {Watanabe},
  \citenamefont {Taniguchi}, \citenamefont {Kulkarni}, \citenamefont {Kuo}
  \emph {et~al.}}]{zhang2020twist}%
  \BibitemOpen
  \bibfield  {author} {\bibinfo {author} {\bibfnamefont {L.}~\bibnamefont
  {Zhang}}, \bibinfo {author} {\bibfnamefont {Z.}~\bibnamefont {Zhang}},
  \bibinfo {author} {\bibfnamefont {F.}~\bibnamefont {Wu}}, \bibinfo {author}
  {\bibfnamefont {D.}~\bibnamefont {Wang}}, \bibinfo {author} {\bibfnamefont
  {R.}~\bibnamefont {Gogna}}, \bibinfo {author} {\bibfnamefont
  {S.}~\bibnamefont {Hou}}, \bibinfo {author} {\bibfnamefont {K.}~\bibnamefont
  {Watanabe}}, \bibinfo {author} {\bibfnamefont {T.}~\bibnamefont {Taniguchi}},
  \bibinfo {author} {\bibfnamefont {K.}~\bibnamefont {Kulkarni}}, \bibinfo
  {author} {\bibfnamefont {T.}~\bibnamefont {Kuo}}, \emph {et~al.},\
  }\href@noop {} {\bibfield  {journal} {\bibinfo  {journal} {Nature
  communications}\ }\textbf {\bibinfo {volume} {11}},\ \bibinfo {pages} {5888}
  (\bibinfo {year} {2020})}\BibitemShut {NoStop}%
\bibitem [{\citenamefont {Tang}\ \emph {et~al.}(2021)\citenamefont {Tang},
  \citenamefont {Gu}, \citenamefont {Liu}, \citenamefont {Watanabe},
  \citenamefont {Taniguchi}, \citenamefont {Hone}, \citenamefont {Mak},\ and\
  \citenamefont {Shan}}]{tang2021tuning}%
  \BibitemOpen
  \bibfield  {author} {\bibinfo {author} {\bibfnamefont {Y.}~\bibnamefont
  {Tang}}, \bibinfo {author} {\bibfnamefont {J.}~\bibnamefont {Gu}}, \bibinfo
  {author} {\bibfnamefont {S.}~\bibnamefont {Liu}}, \bibinfo {author}
  {\bibfnamefont {K.}~\bibnamefont {Watanabe}}, \bibinfo {author}
  {\bibfnamefont {T.}~\bibnamefont {Taniguchi}}, \bibinfo {author}
  {\bibfnamefont {J.}~\bibnamefont {Hone}}, \bibinfo {author} {\bibfnamefont
  {K.~F.}\ \bibnamefont {Mak}},\ and\ \bibinfo {author} {\bibfnamefont
  {J.}~\bibnamefont {Shan}},\ }\href@noop {} {\bibfield  {journal} {\bibinfo
  {journal} {Nature Nanotechnology}\ }\textbf {\bibinfo {volume} {16}},\
  \bibinfo {pages} {52} (\bibinfo {year} {2021})}\BibitemShut {NoStop}%
\bibitem [{\citenamefont {Haastrup}\ \emph {et~al.}(2018)\citenamefont
  {Haastrup}, \citenamefont {Strange}, \citenamefont {Pandey}, \citenamefont
  {Deilmann}, \citenamefont {Schmidt}, \citenamefont {Hinsche}, \citenamefont
  {Gjerding}, \citenamefont {Torelli}, \citenamefont {Larsen}, \citenamefont
  {Riis-Jensen} \emph {et~al.}}]{haastrup2018computational}%
  \BibitemOpen
  \bibfield  {author} {\bibinfo {author} {\bibfnamefont {S.}~\bibnamefont
  {Haastrup}}, \bibinfo {author} {\bibfnamefont {M.}~\bibnamefont {Strange}},
  \bibinfo {author} {\bibfnamefont {M.}~\bibnamefont {Pandey}}, \bibinfo
  {author} {\bibfnamefont {T.}~\bibnamefont {Deilmann}}, \bibinfo {author}
  {\bibfnamefont {P.~S.}\ \bibnamefont {Schmidt}}, \bibinfo {author}
  {\bibfnamefont {N.~F.}\ \bibnamefont {Hinsche}}, \bibinfo {author}
  {\bibfnamefont {M.~N.}\ \bibnamefont {Gjerding}}, \bibinfo {author}
  {\bibfnamefont {D.}~\bibnamefont {Torelli}}, \bibinfo {author} {\bibfnamefont
  {P.~M.}\ \bibnamefont {Larsen}}, \bibinfo {author} {\bibfnamefont {A.~C.}\
  \bibnamefont {Riis-Jensen}}, \emph {et~al.},\ }\href@noop {} {\bibfield
  {journal} {\bibinfo  {journal} {2D Materials}\ }\textbf {\bibinfo {volume}
  {5}},\ \bibinfo {pages} {042002} (\bibinfo {year} {2018})}\BibitemShut
  {NoStop}%
\bibitem [{\citenamefont {Tang}\ \emph {et~al.}(2022)\citenamefont {Tang},
  \citenamefont {Gu}, \citenamefont {Liu}, \citenamefont {Watanabe},
  \citenamefont {Taniguchi}, \citenamefont {Hone}, \citenamefont {Mak},\ and\
  \citenamefont {Shan}}]{tang2022dielectric}%
  \BibitemOpen
  \bibfield  {author} {\bibinfo {author} {\bibfnamefont {Y.}~\bibnamefont
  {Tang}}, \bibinfo {author} {\bibfnamefont {J.}~\bibnamefont {Gu}}, \bibinfo
  {author} {\bibfnamefont {S.}~\bibnamefont {Liu}}, \bibinfo {author}
  {\bibfnamefont {K.}~\bibnamefont {Watanabe}}, \bibinfo {author}
  {\bibfnamefont {T.}~\bibnamefont {Taniguchi}}, \bibinfo {author}
  {\bibfnamefont {J.~C.}\ \bibnamefont {Hone}}, \bibinfo {author}
  {\bibfnamefont {K.~F.}\ \bibnamefont {Mak}},\ and\ \bibinfo {author}
  {\bibfnamefont {J.}~\bibnamefont {Shan}},\ }\href@noop {} {\bibfield
  {journal} {\bibinfo  {journal} {arXiv preprint arXiv:2201.12510}\ } (\bibinfo
  {year} {2022})}\BibitemShut {NoStop}%
\bibitem [{\citenamefont {Koperski}\ \emph {et~al.}(2019)\citenamefont
  {Koperski}, \citenamefont {Molas}, \citenamefont {Arora}, \citenamefont
  {Nogajewski}, \citenamefont {Bartos}, \citenamefont {Wyzula}, \citenamefont
  {Vaclavkova}, \citenamefont {Kossacki},\ and\ \citenamefont
  {Potemski}}]{Koperski2019}%
  \BibitemOpen
  \bibfield  {author} {\bibinfo {author} {\bibfnamefont {M.}~\bibnamefont
  {Koperski}}, \bibinfo {author} {\bibfnamefont {M.~R.}\ \bibnamefont {Molas}},
  \bibinfo {author} {\bibfnamefont {A.}~\bibnamefont {Arora}}, \bibinfo
  {author} {\bibfnamefont {K.}~\bibnamefont {Nogajewski}}, \bibinfo {author}
  {\bibfnamefont {M.}~\bibnamefont {Bartos}}, \bibinfo {author} {\bibfnamefont
  {J.}~\bibnamefont {Wyzula}}, \bibinfo {author} {\bibfnamefont
  {D.}~\bibnamefont {Vaclavkova}}, \bibinfo {author} {\bibfnamefont
  {P.}~\bibnamefont {Kossacki}},\ and\ \bibinfo {author} {\bibfnamefont
  {M.}~\bibnamefont {Potemski}},\ }\bibfield  {journal} {\bibinfo  {journal}
  {2D Materials}\ }\textbf {\bibinfo {volume} {6}},\ \href
  {https://doi.org/10.1088/2053-1583/aae14b} {10.1088/2053-1583/aae14b}
  (\bibinfo {year} {2019})\BibitemShut {NoStop}%
\bibitem [{\citenamefont {Zhang}\ \emph {et~al.}(2021)\citenamefont {Zhang},
  \citenamefont {Shinokita}, \citenamefont {Watanabe}, \citenamefont
  {Taniguchi}, \citenamefont {Miyauchi},\ and\ \citenamefont
  {Matsuda}}]{Zhang2021-Polarization}%
  \BibitemOpen
  \bibfield  {author} {\bibinfo {author} {\bibfnamefont {Y.}~\bibnamefont
  {Zhang}}, \bibinfo {author} {\bibfnamefont {K.}~\bibnamefont {Shinokita}},
  \bibinfo {author} {\bibfnamefont {K.}~\bibnamefont {Watanabe}}, \bibinfo
  {author} {\bibfnamefont {T.}~\bibnamefont {Taniguchi}}, \bibinfo {author}
  {\bibfnamefont {Y.}~\bibnamefont {Miyauchi}},\ and\ \bibinfo {author}
  {\bibfnamefont {K.}~\bibnamefont {Matsuda}},\ }\href
  {https://doi.org/10.1002/ADFM.202006064} {\bibfield  {journal} {\bibinfo
  {journal} {Advanced Functional Materials}\ }\textbf {\bibinfo {volume}
  {31}},\ \bibinfo {pages} {2006064} (\bibinfo {year} {2021})}\BibitemShut
  {NoStop}%
\bibitem [{\citenamefont {Yu}\ \emph {et~al.}(2017)\citenamefont {Yu},
  \citenamefont {Liu}, \citenamefont {Tang}, \citenamefont {Xu},\ and\
  \citenamefont {Yao}}]{yu2017moire}%
  \BibitemOpen
  \bibfield  {author} {\bibinfo {author} {\bibfnamefont {H.}~\bibnamefont
  {Yu}}, \bibinfo {author} {\bibfnamefont {G.-B.}\ \bibnamefont {Liu}},
  \bibinfo {author} {\bibfnamefont {J.}~\bibnamefont {Tang}}, \bibinfo {author}
  {\bibfnamefont {X.}~\bibnamefont {Xu}},\ and\ \bibinfo {author}
  {\bibfnamefont {W.}~\bibnamefont {Yao}},\ }\href@noop {} {\bibfield
  {journal} {\bibinfo  {journal} {Science advances}\ }\textbf {\bibinfo
  {volume} {3}},\ \bibinfo {pages} {e1701696} (\bibinfo {year}
  {2017})}\BibitemShut {NoStop}%
\end{thebibliography}%

\end{document}


\title{Distinctive $g$-factor of moiré-confined excitons in van der Waals heterostructures: Supplementary Information}

\author{Y. Galvão Gobato}\email{yara@df.ufscar.br}

\author{C. Serati de Brito}
\affiliation{Physics Department, Federal University of São Carlos (UFSCAR)-Brazil}

\author{A. Chaves}\email{andrey@fisica.ufc.br}
\affiliation{Universidade Federal do Cear\'a, Departamento de F\'{\i}sica, 60455-760 Fortaleza, Cear\'a, Brazil}
\affiliation{Department of Physics, University of Antwerp, Groenenborgerlaan 171, B-2020 Antwerp, Belgium}

\author{M. A. Prosnikov}
\affiliation{High Field Magnet Laboratory (HFML–EMFL), Radboud University, 6525 ED, Nijmegen, The Netherlands}

\author{T. Wo\'zniak}
\affiliation{Department of Semiconductor Materials Engineering, Wroc\l aw University of Science and Technology, 50-370 Wroc\l aw, Poland}

\author{Shi Guo}
\affiliation{Centre for Graphene Science, College of Engineering, Mathematics and Physical Sciences, University of Exeter, Exeter, EX4 4QF}

\author{Ingrid D. Barcelos}
\affiliation{Brazilian Synchrotron Light Laboratory (LNLS), Brazilian Center for Research in Energy and Materials (CNPEM), 13083-970 Campinas, São Paulo, Brazil}

\author{M. V. Milo\v{s}evi\'c}
\affiliation{Department of Physics, University of Antwerp, Groenenborgerlaan 171, B-2020 Antwerp, Belgium}
\affiliation{NANOlab Center of Excellence, University of Antwerp, Belgium}

\author{F. Withers}
\affiliation{Centre for Graphene Science, College of Engineering, Mathematics and Physical Sciences, University of Exeter, Exeter, EX4 4QF}

\author{P. C. M. Christianen}
\affiliation{High Field Magnet Laboratory (HFML–EMFL), Radboud University, 6525 ED, Nijmegen, The Netherlands}

\maketitle


Figure S1 shows typical PL from different positions at 300K. We observed that in the heterostructure region, the PL peaks are observed at lower energy values at around 1.564 eV and 1.974eV as compared to the MoSe$_2$ and WS$_2$ monolayer regions respectively. Small changes in the PL from different positions are also observed, particularly in the heterostructure region, attributed to some local strain distributions induced by the presence of bubbles in the sample. 
\begin{figure}[!t]
\centering{\includegraphics[width=0.9\columnwidth]{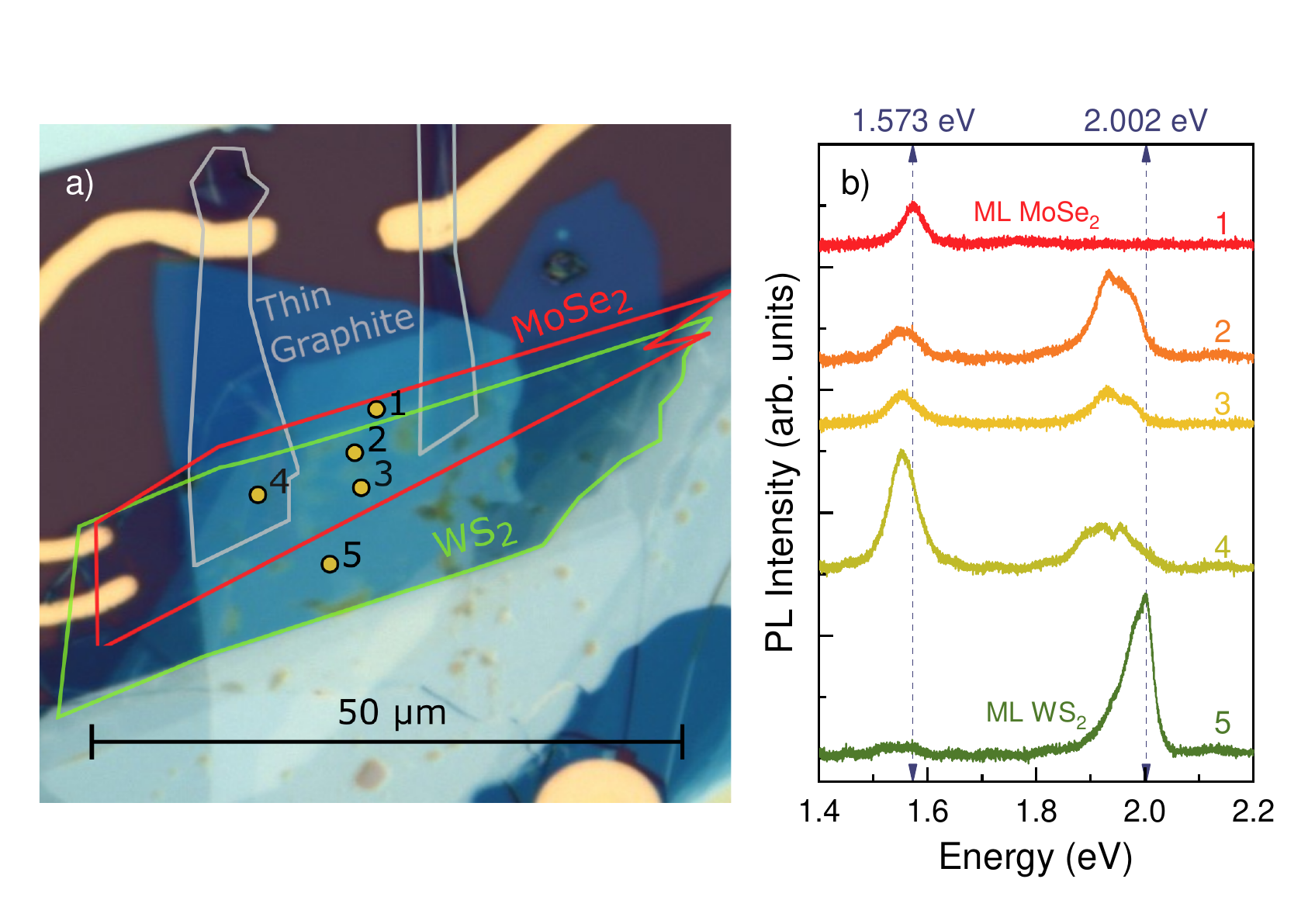}}
\caption{(a) Optical image of the sample and (b) typical PL spectra as measured at different positions, labelled by the numbers in (a), at 300 K.}
\end{figure}

The results in Figs. 2 and 3 of the main manuscript were measured on position labelled P1. Figure S2 shows a typical fitting of PL spectrum at 4 K for a different position, labelled P2. For laser positions close to the bubbles of the sample, such as at P2, we have fitted the intralayer exciton  (X$_{A1}$ and X$_{A2}$) and trion peaks (X$^*_1$ and X$^*_2$) with two lorentzians. Since, in this case, the objective collects PL from regions with different characteristics, namely, on the bubble or around it, the peaks related to the intralayer MoSe$_2$ exciton and trion are split. 

\begin{figure}[!t]
\centering{\includegraphics[width=0.75\columnwidth]{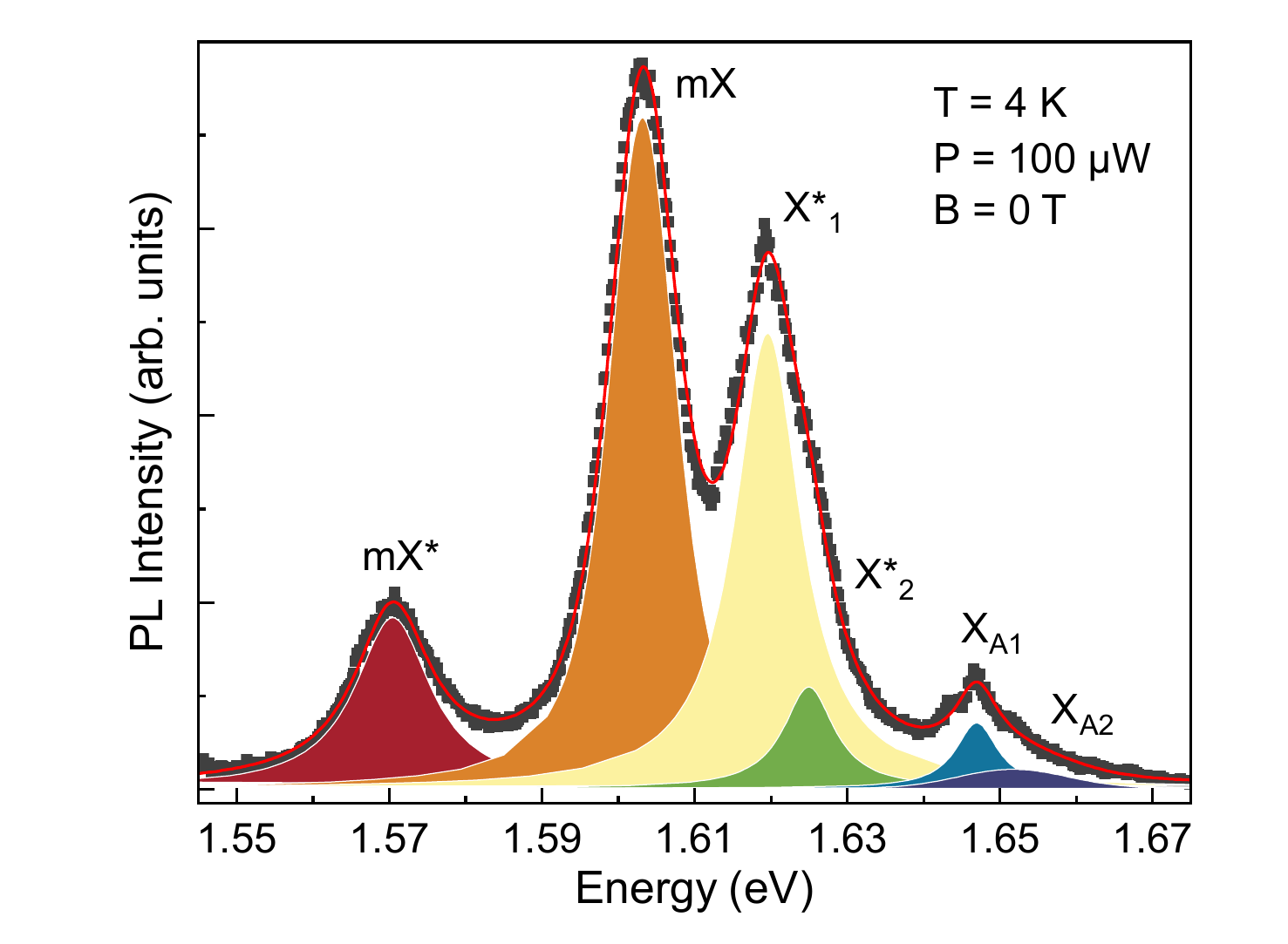}}
\caption{Typical fitting of a PL spectrum measured at position P2, zero magnetic field, temperature $T =$ 4 K, and excitation power $P =$ 100 $\mu$W.}
\end{figure}

\begin{figure}[H]
\centering{\includegraphics[width=1.0\columnwidth]{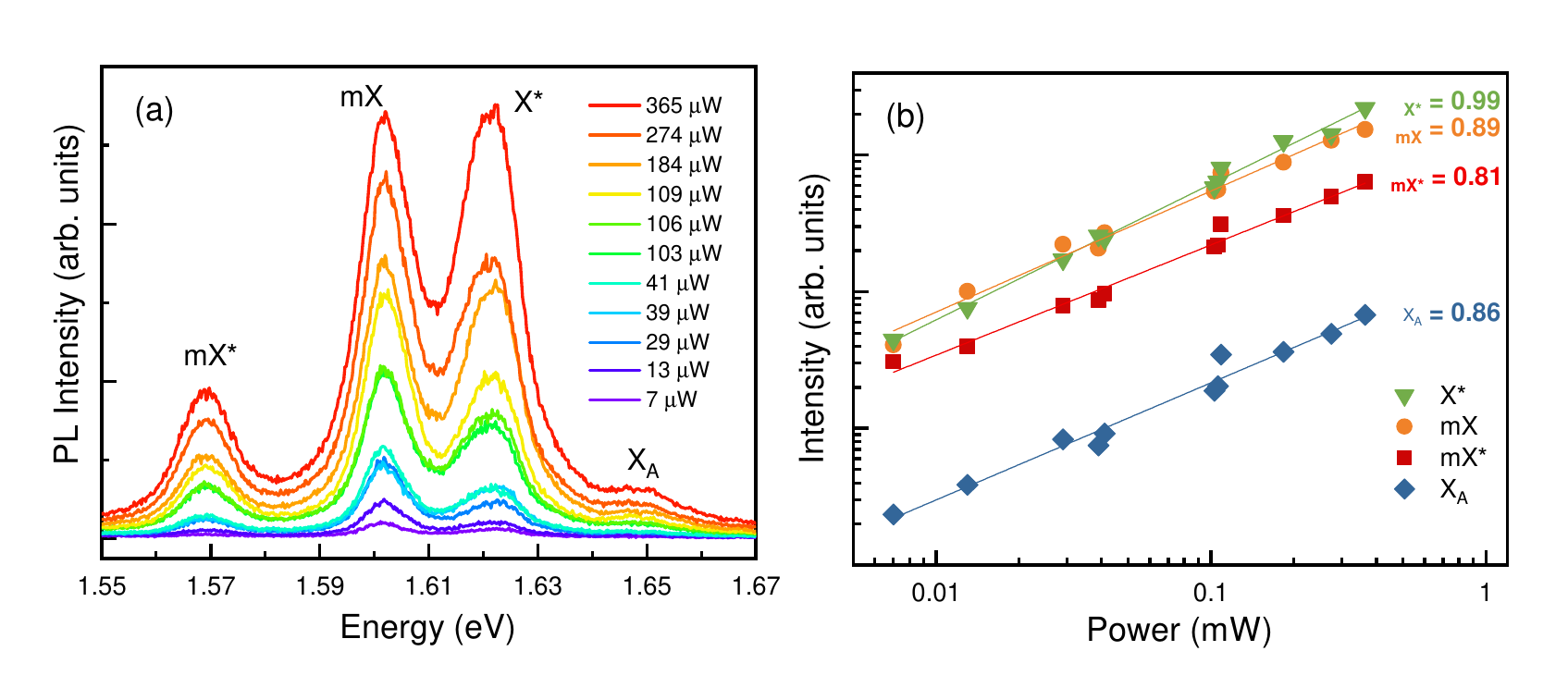}}
\caption{(a) Laser power dependence of the PL spectra at 4 K. (b) Logarithmic plot of  the integrated PL intensity, for all observed PL peaks, as a function of excitation power. Solid curves show $I \approx P^{\alpha}$ fittings, along with their fitting $\alpha$ values.}
\end{figure}

Figure S3(a) shows the laser power dependence of the PL spectra measured in the heterostructure region near position P2. Figure S3(b) shows  the double logarithmic plot of the integrated intensities of the different excitonic peaks as a function of laser power , for the X$^*$, mX, mX$^*$ and X$_A$ peaks. The solid lines are fitting curves $I = P^{\alpha}$, along with their respective $\alpha$ values. Ideally, one expects linear behaviour for excitons, with $\alpha \approx$ 1. We have observed that all peaks exhibit a power-law behaviour with $\alpha$ values $\approx$ 0.8 to 1.0, which confirms the excitonic character of the emission peaks. Small deviations from $\alpha$ $\approx$1 are most likely due to signals captured from other optically induced localized excitons and/or to the fitting procedure precision. 

\begin{figure}[H]
\centering{\includegraphics[width=0.8\columnwidth]{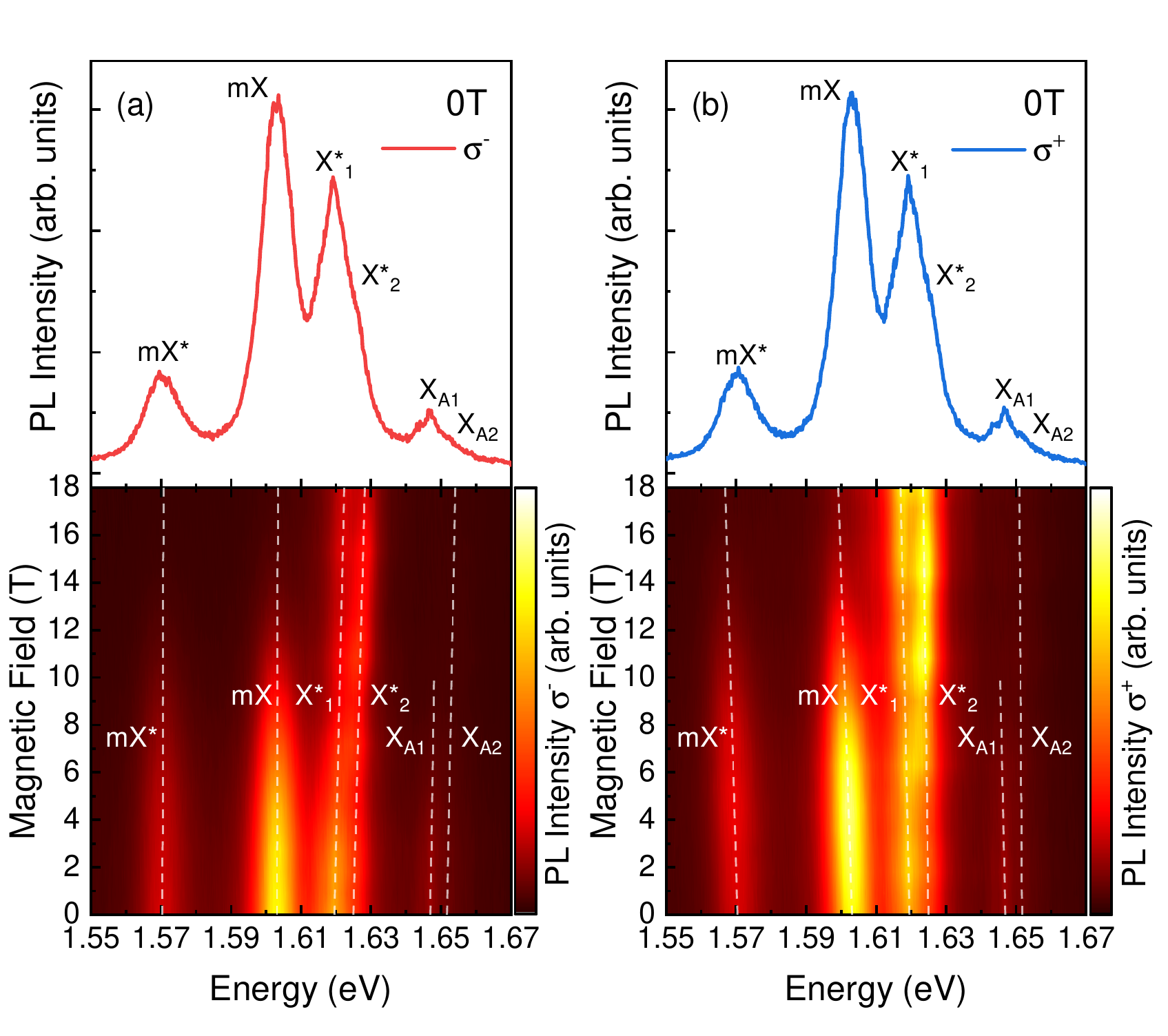}}
\caption{(a) Typical PL spectrum at 0 T and (b) color-code mapping of the magnetic field dependence of $\sigma_+$ and $\sigma_-$ PL of MoSe$_2$/WS$_2$ heterostructures with twisted angle around 0º, measured at temperature 4 K and position P2.}
\end{figure}

Figure S4 shows the magnetic field dependence for the (a) left ($\sigma_-$) and (b) right ($\sigma_+$) circularly polarized PL measured at position P2, under magnetic fields up to 18 T. As previously mentioned, this position corresponds to a strained region (i.e. close to a bubble region) of the heterostructure. We observe similar behaviour as compared to the results measured at other positions (see Fig. 2 of the main manuscript and Fig. S2). However, depending on the laser position, the X$_A$ and X$^*$ bands are very asymmetric, which evidences a contribution of an additional emission peak. This asymmetry is usually observed near bubbles in the sample and can be attributed to non-uniform local strain, which red-shifts the PL peaks to lower energies as compared to those of unstrained regions.

\begin{figure}[H]
\centering{\includegraphics[width=1\columnwidth]{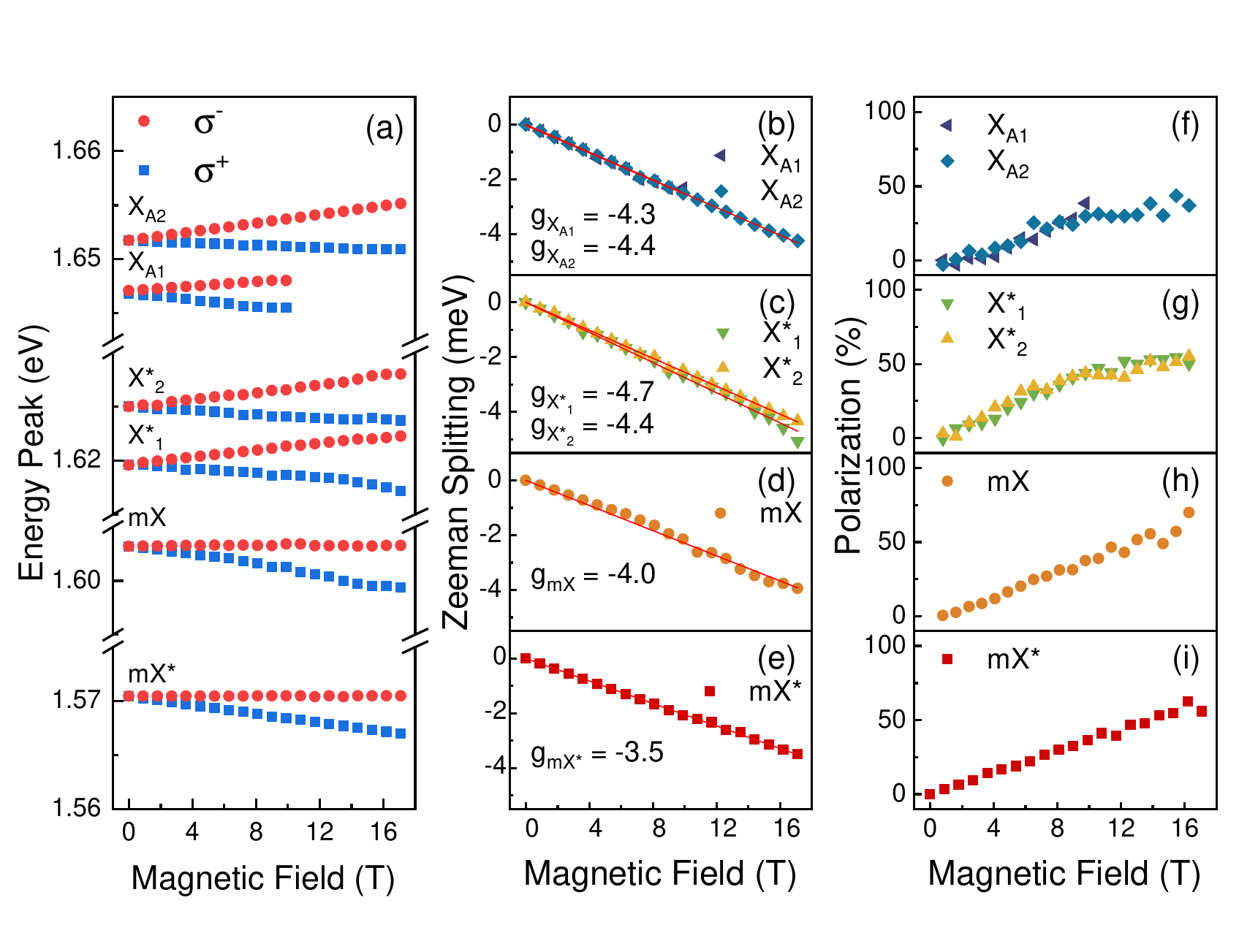}}
\caption{(a) Magnetic field dependence of PL peak energies for $\sigma_+$ and $\sigma_-$ polarization, as measured at position P2. (b)-(e) Zeeman splitting of each PL peak as function of magnetic field (symbols). Solid lines are linear fits to the data. (f)-(i) Polarization degree as a function of the magnetic field for all observed PL peaks.}
\end{figure}

Figure S5(a) shows the PL peak positions versus magnetic field for the peaks identified in the measurements performed at position P2. The effective $g$-factors of X$_{A1}$/X$_{A2}$ and X$^*_1$/X$^*_2$ range from $g \approx$ -4.4 to $\approx$ -4.7, which are similar to (although slightly higher than) the $g$-factors of individual monolayer MoSe$_2$. On the other hand, $g$-factors of the Moiré peaks mX and mX$^*$ (Figs. S5(d) and S5(e)) are  lower, $\approx$ -4.0 and $\approx$ -3.5, respectively. These results are similar to those obtained at position P1, see Fig. 3 of the main text. Figures S5(f)-(g) show the magnetic field dependence of the circular polarization degree for all PL peaks. We have observed a tendency of saturation for X$_{A1}$/X$_{A2}$ and X$^*_1$/X$^*_2$ emission peaks at high magnetic field, while a clear linear behaviour is observed for peaks mX and mX$^*$ and higher values of polarization degree.

We have also performed time resolved PL (TRPL) measurements using a standard TCSPC technique (PicoQuant/Picoharp 300) with a 1.69 eV pulsed laser (PicoQuant / LDH Serie) delivering 70 ps pulses at a repetition frequency of 80 MHz with the signal dispersed by a 75 cm spectrometer and detected by a hybrid PMT device (PicoQuant). Figure S6 shows our  TRPL results. We have observed similar PL decay times  for all  emission peaks. We remark that the PL decay times are very different of typical values  usually obtained for interlayer excitons  in MoSe$_2$/WSe$_2$  heterostructures (around 40ns). Actually, our TRPL results are consistent with Ref.10  which has showed that there is no important hybridization effect for the hX (or mX) in the WS$_2$/MoSe$_2$  heterostructure.

\begin{figure}[H]
\centering{\includegraphics[width=0.75\columnwidth]{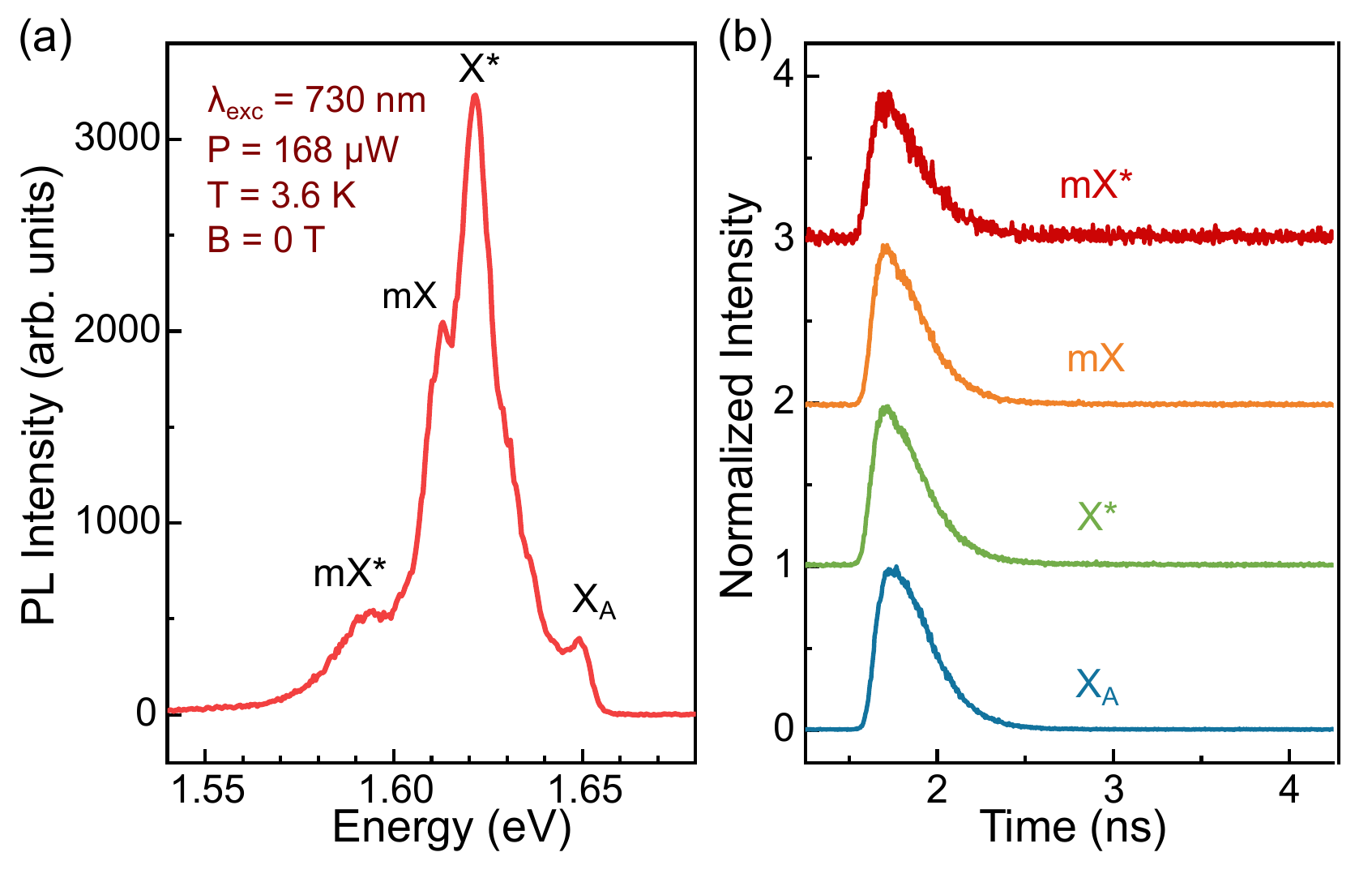}}
\caption{(a) Typical PL spectrum measured with a 1.69 eV pulsed laser for a laser position near P1. (b) PLRT results for all PL peaks. }
\end{figure}

In order to check our theoretical predictions for the $g$-factors in 60$^\circ$ twisted heterostructures, we have also prepared an additional sample with this interlayer twist angle. Although this sample exhibits broader PL peaks, they are similar to the mX and mX$^*$ ones observed in the samples with stacking angle around 0$^\circ$. We have performed magneto-PL measurements for this 60$^\circ$ twisted heterostructure up to lower magnetic fields and extracted the $g$-factor for each emission peak, as shown in Fig. S7.

\begin{figure}[H]
\centering{\includegraphics[width=1\columnwidth]{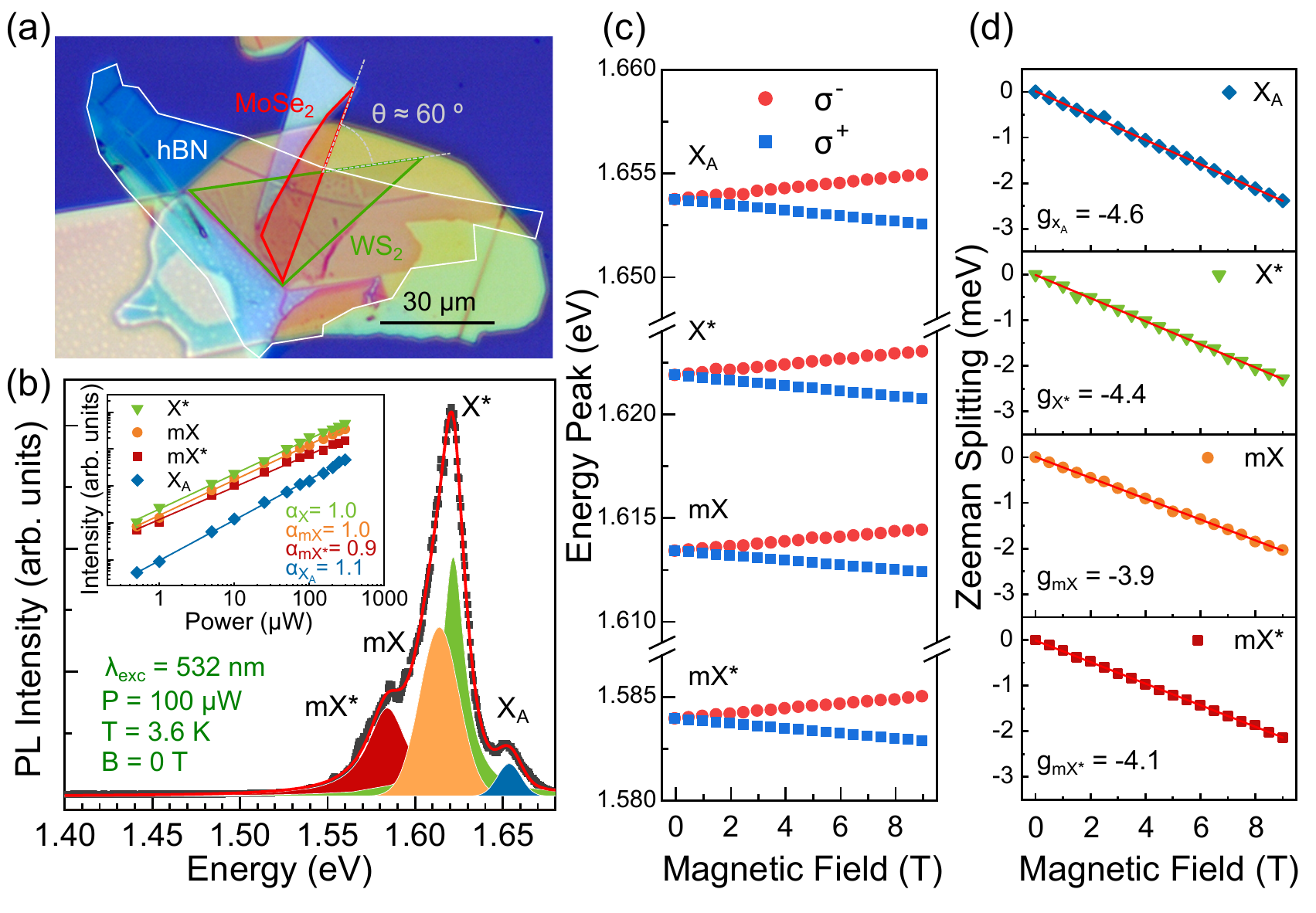}}
\caption{(a) Optical image of a sample of hBN/MoSe2/WS2/hBN with stacking angle around 60$^\circ$. (b) Typical PL spectrum and PL fitting under zero magnetic field at low temperature. (c) Magnetic field dependence of PL peak energies for $\sigma_+$ and $\sigma_-$ polarization for the sample with 60$^\circ$ stacking angle. (d) Zeeman splitting of each PL peak as a function of magnetic field (symbols). Solid lines are linear fits to the data.}
\end{figure}

For the lower magnetic field measurements, we have used a helium closed-cycle cryostat with superconducting magnet coils (Attocube - Attodry1000) with the magnetic field up to 9 T perpendicular to the heterostructure. The sample was mounted on Attocube piezoelectric x-y-z translation stages in order to control the sample position. PL measurements were performed using a continuous-wave (cw) laser with a photon energy of 2.33 eV. The PL signal was collimated using an aspheric lens (NA = 0.64) and the selection of circular polarization components was performed before to be focused into a 50 $\mu$m multimode optical fiber, being dispersed by a 75 cm spectrometer and detected by a silicon CCD detector (Andor, Shamrock/iDus).

Figure S7(a) shows the optical image of the heterostructure with stacking angle around 60$^\circ$. Figure S7 (b) shows the typical PL spectrum at low temperature. The inset shows the double logarithmic plot of the integrated PL intensities of the different excitonic peaks as a function of laser power.  The solid lines are fitting curves $I = P^{\alpha}$, along with their respective $\alpha$ values.  We have observed that all peaks exhibit a power-law behaviour with $\alpha$ values $\approx$ 1.0, which confirms the excitonic character of the emission peaks. Figure S7(c) shows the magnetic field dependence of the Zeeman splitting for all emission peaks. Our results are similar to the results obtained for the sample with stacking angle around 0$^\circ$, i.e., we have observed a small reduction of $g$-factors for mX and mX$^*$, as compared to X$_A$ and T peaks. This result is in agreement with our theoretical predictions for moiré confined excitons with a dominant intralayer component. Finally, we have also performed TRPL measurements (not shown here) for this sample. 0ur results are also similar to the results obtained for sample with stacking angle 0$^\circ$ (Figure S6).

\section{Theoretical model for exciton g-factors}

We have calculated the electronic band structures of MoSe$_2$/WS$_2$ vdWHS with $R_h^h$, $R_h^M$ and $R_h^X$ stacking registries using density functional theory (DFT) with the Vienna Ab Initio Simulation Package (VASP) \cite{kresse1996efficient}, using Perdew-Burke-Ernzerhof (PBE) \cite{perdew1996generalized} and Heyd-Scuseria-Ernzerhof (HSE06) \cite{krukau2006influence} functionals, along with the Projector Augmented Waves technique. \cite{kresse1999ultrasoft,blochl1994projector} All unit cells were built with at least 15 \AA\, separation between replicates in the perpendicular direction to achieve negligible interlayer interaction. A 500 eV cut-off energy for plane wave basis and 12$\times$12$\times$1 $\Gamma$-centered Monkhorst-Pack $k$-grid for Brillouin zone sampling were used. Spin orbit coupling was taken into account during all the calculations. Interlayer van der Waals interactions were considered via semi-empirical D3 correction of Grimme. \cite{grimme2010consistent} 

For monolayer(1L)-MoSe$_2$, atomic positions and lattice constants were optimized with 10$^{-3}$  eV/\AA\, and 0.1 kbar precision. For vDWHSs, only atomic positions were optimized, while the lattice constants were kept fixed at the 1L-MoSe$_2$ value. This results in +3.6\% strain for WS$_2$ layer, which artificially downshifts its valence and conduction bands. This, however, does not influence the Bloch states of the MoSe$_2$ layer, which are considered in further analysis of e.g. angular momentum calculations.

Table \ref{tab:1} provides the energy gaps and effective masses obtained by our DFT calculations. Results for 1L-MoSe$_2$ are also shown for comparison. Since the energy gap in $R_h^X$ stacking is lower than that for other stacking registries, this gap is expected to be locally lower in the regions of the moiré pattern that can be identified as (approximately) $R_h^X$, which would then confine the exciton wave function to those regions. 

\begin{table}
\caption{Direct energy gaps $E_g$ (in eV) at K-point, obtained with PBE and HSE06 functionals, and electron (hole) effective masses $m_{e(h)}$ (in units of $m_0$) in 1L-MoSe$_2$, as well as in MoSe$_2$/WS$_2$ vdWHS with $R_h^h$, $R_h^M$ and $R_h^X$ stacking registries.} 
\label{tab:1}\renewcommand{\arraystretch}{1.5}

\begin{tabular}{c  c c  c c }
\hline
 & $E_g (PBE)$ & $E_g (HSE06)$ & $m_{h}$ ~~ & $m_e$\\
\hline
1L-MoSe$_2$  &  1.411 &   1.830 &  0.582 ~~ & 0.512 \\
$R_h^h$  &  1.410 &   1.830 &  0.580 ~~ & 0.514  \\
$R_h^X$  &  1.400 &   1.816 &  0.578 ~~ & 0.516 \\ 
$R_h^M$  &  1.411 &   1.831 &  0.585 ~~ & 0.515 \\
\hline  
\end{tabular}

\end{table}

With the obtained energy bands $\epsilon_{n\textbf{k}}$ and Bloch states $|n\textbf{k}\rangle$, we calculate the effective $g$-factor of conduction and valence band states as 
\begin{equation}
g_{n\textbf{k}} = L_{n\textbf{k}} + \Sigma_{n\textbf{k}}, \end{equation}
where $\Sigma_{n\textbf{k}} = \langle n\textbf{k} | \Sigma^{z} |m \textbf{k}\rangle$ accounts for the spin angular momentum in the direction perpendicular to the material plane, while the z component of orbital angular momentum matrix elements are evaluated as
\begin{equation}
L_{n\textbf{k}} = -\frac{i}{m_0}\sum_{n,m\neq n}\frac{p^{x}_{nm\textbf{k}}p^{y}_{mn\textbf{k}} - p^{y}_{nm\textbf{k}}p^{x}_{mn\textbf{k}}}{\epsilon_{n\textbf{k}}-\epsilon_{m\textbf{k}}},     
\end{equation}
with momentum operator matrix elements $p^{\alpha}_{nm\textbf{k}} = \langle n\textbf{k} | p^{\alpha} |m \textbf{k}\rangle$. For details on the derivation of these expressions, refer to Ref. [\onlinecite{wozniak2020exciton}], where values for 1L-MoS$_2$, 1L-WS$_2$, 1L-MoSe$_2$, 1L-WSe$_2$ and MoSe$_2$/WSe$_2$ vdWHS are also provided. Exciton $g$-factors are then estimated from the linear dependence of the quasi-particle energy gap on the magnetic field $E_k (B) = (g_{c\textbf{k}}- g_{v\textbf{k}})\mu_B B = g_i \mu_B$, where $i = $ A (free exciton) or mX (moiré confined exciton), as defined in the main text, and $g_{c(v)\textbf{k}}$ is the $g$-factor of the conduction (valence) band state involved in each exciton state.  

All vdWHSs exhibit electronic states consistent with type-I band alignment (both conduction and valence band edge states mostly in the MoSe$_2$ layer), with practically no hybridization, i.e. no Bloch wave functions shared between the MoSe$_2$ and WS$_2$ layers. This agrees both with previous DFT calculations for other TMD-based vdWHS \cite{gao2017interlayer, chaves2018electrical}, and with experimental results \cite{tang2021tuning}, and is a consequence of the fact that the K-point states in the individual monolayers mostly involve $d$-orbitals of the transition metal atoms, which are buried in between the chalcogens, therefore, they are not expected to exhibit strong coupling as the two monolayers are stacked together in a vdWHS. Nevertheless, it is still important to discuss the consequences of hybridization with an inter-layer state, as proposed in Refs. [\onlinecite{alexeev2019resonantly,zhang2020twist}], since such hybridization may also be achieved and controlled even by perpendicularly applied electric fields in any vdWHS \cite{gao2017interlayer, chaves2018electrical, tang2021tuning}. Besides, in Ref. [\onlinecite{kunstmann2018momentum}], experimental results suggest a hybridized ground-state exciton with a hole distributed over both layers in MoS$_2$/WSe$_2$ vdWHS. 

By applying a vertical bias, electrons may be pushed towards either the MoSe$_2$ or the WS$_2$ layer, which would change the electronic distribution between the layers \cite{tang2021tuning}. Since the angular momenta of the conduction band states in each of these layers are strikingly different \cite{wozniak2020exciton}, the effective exciton $g$-factor $(g_{c\textbf{k}}-g_{v\textbf{k}})$ will also be strongly controlled by the bias, varying from $g \approx -4$ for a pure intra-layer state (as in our experimental data) to $g \approx -2$ for an inter-layer exciton induced by strong bias.  The experimental evidence of such bias-dependent $g$-factors is a worthy outlook for future studies. Furthermore, the control of strain and twist angles could also provide interesting ways to modify $g$-factors of hybridized excitons. In the former, strain can be used to modify the conduction band edges for different materials in different amounts, thereby modifying the conduction band offset and, consequently, affecting the inter-layer hybridization of the electronic wave function. Qualitatively, the discussion presented here is relevant in all these cases, where $g$-factors of hybridized states are expected to be either higher or lower than those of their intra-layer excitons, depending on their material compositions, which can be predicted using e.g. the angular momenta of conduction and valence band states as calculated in Ref. [\onlinecite{wozniak2020exciton}].

\bibliographystyle{apsrev4-2}
\bibliography{references}